\documentclass[smallextended]{svjour3}       
\smartqed  
\usepackage[sort]{natbib}
\usepackage[table]{xcolor}
\usepackage{amsmath,amssymb,amsfonts}
\usepackage{algorithmic}
\usepackage{graphicx}
\usepackage{enumitem}
\usepackage{textcomp}
\usepackage[T1]{fontenc}
\usepackage{listings}
\usepackage{float}
\usepackage[font=small,skip=0pt]{caption}
\usepackage[linesnumbered,ruled]{algorithm2e}
\usepackage{threeparttable}
\usepackage{subcaption}
\usepackage{examplep}
\usepackage{booktabs}
\usepackage{pifont}
\usepackage{setspace}
\usepackage{cprotect}
\usepackage{todonotes}
\usepackage{framed}
\usepackage{tcolorbox}
\usepackage{flushend}
\usepackage{multirow}
\usepackage{wrapfig}

\SetKwProg{Fn}{Function}{}{end}	
\SetAlFnt{\small}
\SetAlCapFnt{\small}
\SetAlCapNameFnt{\small}
\SetKwBlock{Foreach}{for each}{end}
\SetCommentSty{slfamily}

\appto\TPTnoteSettings{\footnotesize}   

\lstdefinestyle{mystyle}{%
  frame            = tblr,    
  tabsize          = 1,     
  numbers          = left,  
  numbersep = 4pt,  
  framesep         = 3pt,   
  framerule        = 0pt, 
  commentstyle     = \color{darkgray},      
  keywordstyle     = \color{blue},       
  showstringspaces = false,              
  escapeinside={<@}{@>}
}

\begin{document}

\title{ConfigCrusher: Towards White-Box Performance Analysis for Configurable Systems}


\author{Miguel Velez 
             \and
            Pooyan Jamshidi
            \and
            Florian Sattler
            \and
            Norbert Siegmund
            \and
            Sven Apel
            \and
            Christian K{\"a}stner
}


\institute{Miguel Velez \at
               Carnegie Mellon University
               \and
               Pooyan Jamshidi \at
			 University of South Carolina
			 \and
			 Florian Sattler \at
			 Saarland University
			 \and
			 Norbert Siegmund \at
			 Leipzig University
			 \and
			 Sven Apel \at
			 Saarland University, Saarland Informatics Campus
			 \and
			 Christian K{\"a}stner \at
			 Carnegie Mellon University
}

\date{Received: date / Accepted: date}

\maketitle

\begin{abstract}
Stakeholders of configurable systems are often interested in knowing how configuration options influence the performance of a system to facilitate, for example, the debugging and optimization processes of these systems.
Several black-box approaches can be used to obtain this information, but they either sample a large number of configurations to make accurate predictions or miss important performance-influencing interactions when sampling few configurations.
Furthermore, black-box approaches cannot pinpoint the parts of a system that are responsible for performance differences among configurations.
This article proposes {\sf ConfigCrusher}, a white-box performance analysis that inspects the implementation of a system to guide the performance analysis, exploiting several insights of configurable systems in the process.
{\sf ConfigCrusher} employs a static data-flow analysis to identify how configuration options may influence control-flow statements and instruments code regions, corresponding to these statements, to dynamically analyze the influence of configuration options on the regions' performance.
Our evaluation on $10$ configurable systems shows the feasibility of our white-box approach to more efficiently build performance-influence models that are similar to or more accurate than current state of the art approaches.
Overall, we showcase the benefits of white-box performance analyses and their potential to outperform black-box approaches and provide additional information for analyzing configurable systems.
\end{abstract}

\section{Introduction}
\label{intro}

Most of today's software systems, such as databases, Web servers, processing libraries, and compilers, provide configuration options to satisfy a large variety of requirements~\citep{SGAK:ESECFSE15, ABKS:FOSPL13, JVKSK:SEAMS17, XJFZPT:ESECFSE15, KSKGA:SOSYM18}. To this end, stakeholders select specific values for each option to obtain the desired functional properties and quality attributes in the system.
However, this configuration process is often a difficult task, especially when lacking knowledge of how the configuration options influence the functionality and qualities of the system~\citep{ABKS:FOSPL13, SGAK:ESECFSE15, XZHZSYZP:SOSP13}.
For this reason, users, developers, and administrations typically resort to default configurations or change individual options in a trial-and-error fashion without understanding the resulting effect~\citep{XJFZPT:ESECFSE15, JC:MASCOTS16, JQCR:CICSE14, HXC:VAMOS12}.\looseness=-1

Performance is one of the many interesting qualities of such systems. 
Understanding how individual configuration options and their combinations influence the performance of the system would facilitate the reasoning, debugging, adaptation, and optimization processes of these systems~\citep{HYL:ASE18, KSKGA:SOSYM18, WLHLSK:ASPLOS18, ZLGBMLSY:SOCC17, SGAK:ESECFSE15, XZHZSYZP:SOSP13, HY:ESEM16}. For example, users can find the configuration that performs an execution the fastest, and developers can find configuration options that cause excessive execution time when debugging the system.

\begin{figure}[t]
\begin{center}
\includegraphics[width=0.45\textwidth]{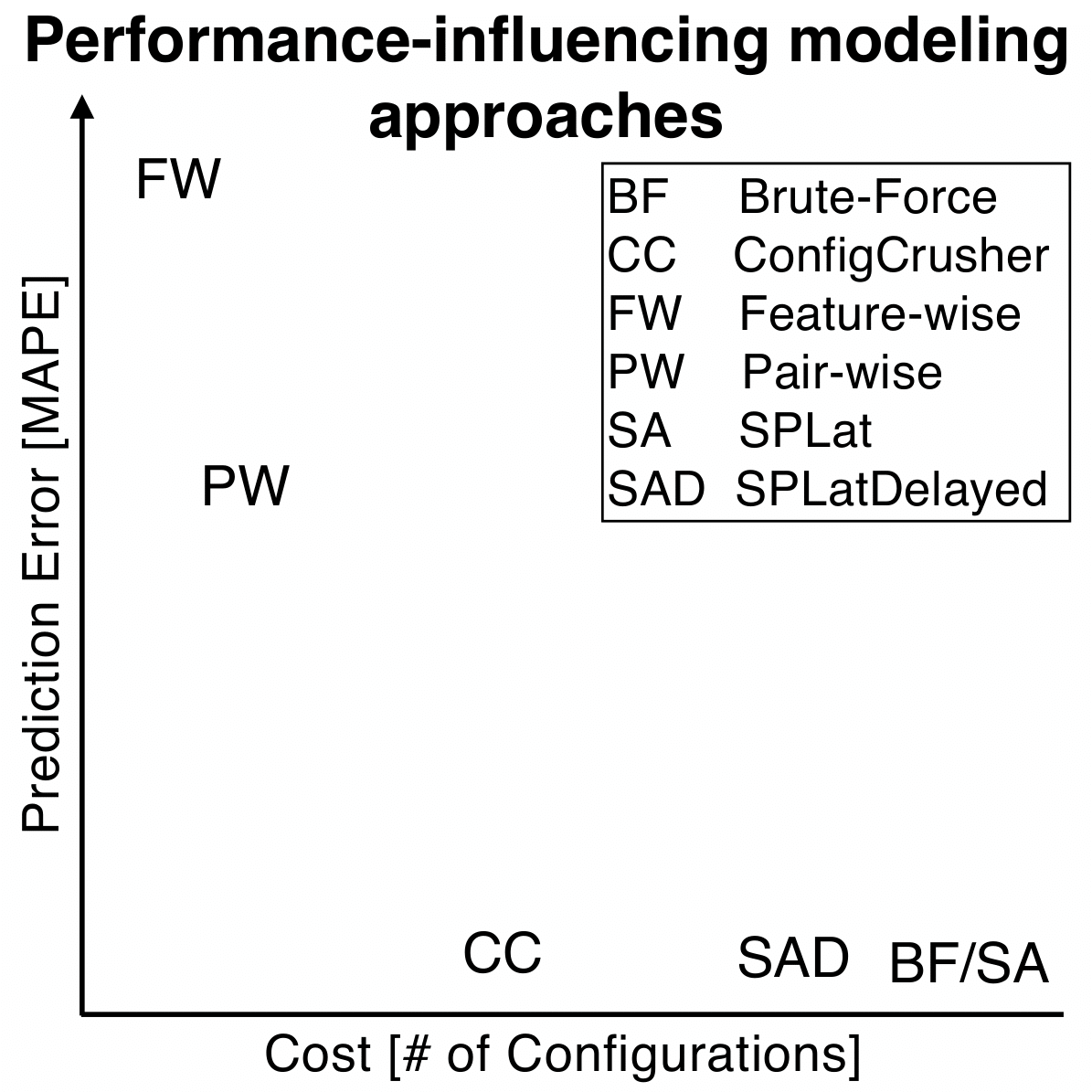}
\end{center}
\caption{Our conjecture on cost and prediction error comparison between state of the art approaches.}
\label{concept-cost-error}
\end{figure}

One research area has focused on understanding the influence of options and their interactions on the performance of a configurable system by building a \emph{performance-influence model}~\citep{SGAK:ESECFSE15}, which describes the performance of a system in terms of its configuration options for a specific workload and in a specific environment (e.g., on given hardware).
Most prior work on deriving performance-influence models uses \emph{black-box approaches} \citep{SRKKAS:SQJ12, SGAK:ESECFSE15, HHL:LION11, ORGC:SPLC14}, which consider the system as a black box and measure for how long the system executes in different configurations.
These approaches sample a subset of the configurations of a system and extrapolate a model based on the corresponding end-to-end measurements.
The model's accuracy and cost depend on the approaches' sampling strategy (i.e., which configurations to measure) and the algorithm used for learning~\citep{KSKGA:SOSYM18}.
Sampling is particularly important: The accuracy of a model might be low if the sample set 
does not capture performance-influencing interactions among options.
Several different sampling strategies with different cost-accuracy tradeoffs have
been explored (cf.\ Fig.~\ref{concept-cost-error}).\looseness=-1

We argue that a \emph{white-box approach} that analyzes the system's source code can 
provide additional insights and guide the performance analysis to relevant options and
interactions.
Where black-box approaches are blind to the internals of an implementation, white-box
approaches can, for example, identify options interacting in control-flow statements,
and thus focus measurements on fewer, but more relevant configurations,
thus promising to build more accurate models at lower cost (cf.\ Fig.~\ref{concept-cost-error}).
In addition, analyzing the implementation and performing measurements with regard
to regions of the system rather than black-box end-to-end measurements allows
us to pinpoint which regions of a system (e.g., set of statements influenced by the same set of options) are responsible for performance differences 
among configurations (i.e., more informative models), which can further help 
in performance debugging and optimization.\looseness=-1


In this article, we introduce a novel white-box performance analysis approach for configurable systems, named {\sf ConfigCrusher}.
It combines static and dynamic analyses to identify and efficiently measure configurations that are relevant for accurate performance modeling.
Specifically, {\sf ConfigCrusher} (1)~uses \emph{static data-flow analysis} to trace the effect that configuration options may directly or indirectly have on a system's control-flow statements (including loops) and 
(2)~\emph{instruments} the system at configuration-relevant control-flow statements 
to measure the performance per region in concrete executions in a small set of selected configurations. 
One key benefit of measuring performance per region is that, in a single executed configuration, we independently measure the influence of multiple options on different regions, a process we call \emph{compression}. 
Overall, {\sf ConfigCrusher} exploits multiple insights about configurable systems, established in several prior studies~\citep{KMKBSBD:ESECFSE13, JSVKPA:ASE17, KBK:AOSD11, MWKTS:ASE16, SKKABRS:ICSE12, SRA:GPCE13, KSKGA:SOSYM18, LKB:TSE18, RSMFP:ICSE10, NKCFP:FSE16}:
\begin{enumerate}[label=(\alph*),noitemsep,nolistsep,topsep=0pt]
 \item \emph{Irrelevance}: Not all options influence the performance of a system on a given workload. {\sf ConfigCrusher}'s data-flow analysis identifies options that do not influence the execution, reducing the number of configurations to sample.\looseness=-1
 \item \emph{Orthogonality}: Not all options interact with each other. {\sf ConfigCrusher}'s data-flow analysis identifies options that are orthogonal and can thus be measured together in a single execution, reducing the number of configurations to sample.
 \item \emph{Low Interaction Degree}: Considering interactions is essential for accurate performance-influence models, but most options tend to interact only with few other options. {\sf ConfigCrusher}'s analysis identifies which interactions can occur, focusing the sampling towards performance-relevant configurations.
\end{enumerate}
Compared to state of the art black-box approaches, {\sf ConfigCrusher} \emph{reduces the cost} of performance modeling while \emph{preserving or increasing the accuracy} of the resulting models.
Guided by program analysis, it will often measure fewer performance-relevant configurations, and, with instrumentation and compression, each measurement can provide information about multiple options and interactions.
Furthermore, in addition to traditional performance-influence models, {\sf ConfigCrusher} builds a \emph{local performance-influence model} for each region of a system.
These models can provide additional fine-grained information and insights to developers, for enhanced debugging and understanding of the performance behavior of a system.

To demonstrate the potential of our white-box approach, we implemented {\sf ConfigCrusher} for Java systems.
using the static taint-analysis engine \emph{FlowDroid}~\citep{ARFBBKLOM:PLDI14} for tracking configuration options.
We show that our white-box approach outperforms 
existing state of the art approaches (i.e., more accurate models at lower cost) 
on $10$ configurable systems.
Due to known scalability issues with precise and accurate static data-flow analyses~\citep{AKKGZARB:ICSE15, ARFBBKLOM:PLDI14, B:ICSENIER18, LSBM:ASE15, DALBSM:ISSTA17, QWR:ISSTA18, WZR:SOAP16, PBW:FSE18}, 
we limit our evaluation to relatively small, but still real-world open-source applications from different domains, including command-line programs, processing libraries, databases, and software product lines.
Nevertheless, our evaluation still provides evidence for the feasibility of the underlying ideas.
More generally, we show the potential benefits of analyzing the system structure to help in the understanding, debugging, and optimization of configurable systems and provide a foundation for future research to scale white-box performance analyses.\looseness=-1

In summary, we make the following contributions:
\begin{itemize}[noitemsep,topsep=0pt]
\item A white-box program analysis, combining data-flow and control-flow analysis and dynamic instrumentation for fine-grained performance measurement to identify how options affect the execution of configurable systems, exploiting insights about common characteristics of such systems~(Sec~\ref{approach}).\looseness=-1
\item A compression technique that allows us to independently infer the influence of options and their interactions on multiple regions of a system's execution and a corresponding method to build accurate performance-influence models~(Sec~\ref{compression}).\looseness=-1
 \item An optimization to reduce the overhead of the instrumented systems that we analyze~(Sec~\ref{instrument}).\looseness=-1
 \item An empirical evaluation of {\sf ConfigCrusher} on 10 systems demonstrating the feasibility and potential of a white-box approach to reduce the cost and increase the accuracy of performance modeling compared to state of the art performance modeling approaches~(Sec~\ref{evaluation}).\looseness=-1
 \item A public open-source implementation of {\sf ConfigCrusher} and reimplementations and improvements of prior state of the art approaches for Java systems~\citep{VJSSAK:ASEJ20SM}. 
 \item A replication package with technical information of the systems analyzed, environmental setup for experiments, analysis scripts, and data of several months of measurements~\citep{VJSSAK:ASEJ20SM}.
\end{itemize}

\section{Performance Modeling of Configurable Systems}
\label{state-of-the-art}
A system's performance and, often directly correlated, energy consumption are
important concerns for many software systems.
While it is possible to design and optimize a system's implementation for a specific task,
much of our modern software is built on top of reusable (open-source) infrastructure software,
such as databases, Web servers, video encoders, and so forth.
However, a single reusable infrastructure component will rarely ever satisfy all stakeholders
equally (``no one size fits all''); for example, developers needing a data-storage
component for a write-heavy scenario would likely make different implementation decisions than 
for a read-heavy scenario, if they would implement such component from scratch.
To resolve this tension between a single reusable component and custom implementations
for different stakeholders, reusable components often have a large number of
configuration options that defer design decisions and allow users to choose between different implementations
and different resulting functionality and quality tradeoffs to meet their needs (e.g., workloads and environments).

When reusing such infrastructure components with many options, making suitable configuration decisions can be challenging~\citep{XJFZPT:ESECFSE15, ABKS:FOSPL13}.
Users are often unaware of the impact on options on various qualities
or have only vague intuitions~\citep{XZHZSYZP:SOSP13, HXC:VAMOS12},
and options may interact, producing surprising behavior~\citep{SGAK:ESECFSE15}.
For example, a developer considering whether to enable encryption, would likely expect that encryption may
slow down system performance, but may need to perform experiments
to identify the severity of the effect, given the concrete system, workload, and other
configuration decisions.

In this context, \emph{performance-influence models} that explain
how configuration options and their interactions influence the performance
of a system, in a certain context (e.g., requirements and needs), can be helpful when making deliberate configuration decisions (e.g., optimizing performance for given workload or debugging the system's
performance behavior).
That is, performance-influence models have a \emph{fundamentally different
approach and goal} than traditional performance models, which typically
model and analyze (e.g., using Queuing networks, Petri Nets, and Stochastic Process Algebras) 
the performance of a system's architecture under different workloads
in the design stage of a project~\citep{H:PMDCSQTA13, SCBSCB:QEST06}, possibly also modeling design decisions as configuration options \citep{EEM:TSE13, BKR:JSS09}.
In contrast, performance-influence models describe the performance behavior 
\emph{of a given system implementation, with a given workload and environment},
in terms of configuration decisions and they are typically \emph{learned
from observing a specific system execution} under different configurations. 



Performance-influence models are typically learned by fitting a model to explain the performance in terms of configuration options~\citep{SGAK:ESECFSE15, GCASW:ASE13,JSVKPA:ASE17, JVKS:FSE18, VPGFC:ICPE17}.
The models can be used for performance debugging~\citep{SGAK:ESECFSE15, WLHLSK:ASPLOS18, XZHZSYZP:SOSP13}, optimization~\citep{OBMS:ESECFSE17, ZLGBMLSY:SOCC17, GCASW:ASE13}, and adaptation~\citep{WLHLSK:ASPLOS18, ZLGBMLSY:SOCC17, JVKSK:SEAMS17, JVKS:FSE18}.

As a running example, the performance-influence model $\Pi = 1 + 3\texttt{A} + 3\texttt{AB} + 3\texttt{AC}$
for the system in Figure~\ref{running-example} (Lines~$1$--$15$) suggests that the measured execution, with a given workload and environment, takes $1$ second by default (Line~\ref{l:exec1}), but $3$ seconds longer if option \texttt{A} (compression) is selected (Lines~\ref{r:influenceac2} $+$ \ref{t:a})
and another $3$ seconds extra, each, if option \texttt{B} (page-size, Line~\ref{r:unignore2}) or \texttt{C}  (encryption, Line~\ref{r:influenceac1}) are selected together
with \texttt{A}.
Such model can help with several maintenance and understanding tasks.
For example, users can optimize the execution time (e.g., deselect \texttt{A} in this example)
or make an informed tradeoff decision, (e.g., whether the functionality of \texttt{A} is
worth the $3$ second overhead).
Likewise, such models can generally help with system understanding
such as identifying the interaction of \texttt{A} with \texttt{B} and \texttt{C}
or recognizing that \texttt{A} might be a performance bottleneck.
Similarly, we can use this model in a planning algorithm to adapt (i.e., dynamically reconfigure) the system, 
in response to changing requirements or environment conditions, such as
high load or low battery~\citep{AGKLMRSSTVVBGHCJ:IEEES19, WLKS:SEAMS17}.

Performance-influence modeling is different from \emph{performance tuning}
approaches that attempt to change a system's configuration to optimize
performance (possibly also considering multiple qualities with a suitable fitness function)
for a given workload and environment~\citep{ORGC:SPLC14, HHL:LION11}.
These approaches search for a good configuration rather than building models of the entire configuration space.
Performance tuning approaches for configurable
systems conduct some sort of search in the configuration space, with and without building intermediate models to support the search~\citep{JC:MASCOTS16, OBMS:ESECFSE17, SRKKAS:SQJ12},
typically measuring the system execution under different configurations.
For performance tuning, search is typically more efficient than building performance-influence
models, because the focus is only on finding the fastest configuration,
not on explaining why it is fast, not modeling the performance of slower configurations,
and not characterizing the influence of options and interactions.
In contrast, performance-influence models, the focus of this article, are more suited for tasks related to understanding,
debugging, prediction, adaptation, and automated reasoning.

\subsection{State of the art of building performance-influence models}

\begin{figure}[t]
\lstset{style=mystyle}
\begin{lstlisting}[language=Java,
morekeywords={def},
						 tabsize = 2,
						 numbers = left, 
						 % escapechar = ~, 
						 basicstyle = \ttfamily\footnotesize, 
						 % linewidth = .6\linewidth,
						 mathescape = true,
						 firstnumber = auto,
						 name = wordpress,
						 xleftmargin = 2em,
						 framexleftmargin=1.5em,
						 ]
def foo(boolean x)    $\label{r:unignore1}$ $\label{p:whole1}$
    // <@\textcolor{red}{Begin region R$_{1}$}@>
    if(x) ... // execution: 4s       <@$\Pi_{R_{1}}$ = 1A + 3AC@> $\label{r:influenceac1}$
    else ... // execution 1s $\label{r:influenceac2}$
    // <@\textcolor{red}{End region R$_{1}$}@>
def main(List workload) $\label{r:noinfluence1}$
    a = getOpt("A"); b = getOpt("B"); $\label{r:loadoptions1}$
    c = getOpt("C"); d = getOpt("D"); $\label{r:loadoptions2}$
    e = getOpt("E"); f = getOpt("F"); $\label{r:loadoptions3}$
    g = getOpt("G"); h = getOpt("H"); $\label{r:loadoptions4}$
    i = getOpt("I"); j = getOpt("J"); $\label{r:loadoptions5}$
    ... // execution: 1s $\label{l:exec1}$
    boolean x = false; $\label{r:noinfluence2}$
    // <@\textcolor{red}{Begin region R$_{2}$}@>
    if(a) // variable depends on option A $\label{r:ifa}$ $\label{r:influencea1}$
        ... // execution: 2s   $\label{t:a}$
        foo(c); // variable depends on option B
        x = true;                    <@$\Pi_{R_{2}}$ = 2A@> $\label{r:influencea2}$
    // <@\textcolor{red}{End region R$_{2}$}@>
    // <@\textcolor{red}{Begin region R$_{3}$}@>
    if(b && x) ... // execution: 3s  <@$\Pi_{R_{3}}$ = 3AB@> $\label{r:unignore2}$ $\label{r:influenceab}$
    // <@\textcolor{red}{End region R$_{3}$}@>
    if(d && e && f) ... // execution: 5s $\label{r:ignore1}$
    if(a) ... // execution: 0.1s  
    if(b) ... // execution: 0.2s  
    if(c) ... // execution: 0.3s  
    if(d) ... // execution: 0.4s  
    if(e) ... // execution: 0.5s  
    if(f) ... // execution: 0.6s  
    if(g) ... // execution: 0.7s  
    if(h) ... // execution: 0.8s  
    if(i) ... // execution: 0.9s  $\label{r:ignore2}$ $\label{p:whole2}$
\end{lstlisting}
\caption{Running example system with three regions, indicated between red comments, influenced by configuration options. The comments indicate the execution time of the branch of each control-flow statement for a given workload, input size, and underlying hardware. For simplicity, we ignore the regions in Lines~\ref{r:ignore1}--\ref{r:ignore2} through Sec.~\ref{implementation}. Region~$1$ is influenced by a control-flow interaction and Region~$3$ by a data-flow interaction. The local performance-influence models are shown to the right of each region.}
\label{running-example}
\end{figure}

Performance-influence models are typically created by selecting a set of 
configurations, measuring the performance of a system for each configuration
(given a fixed workload and environment), and then fitting a model (e.g., a linear model)
to explain the system's performance behavior in terms of its configuration options.
The accuracy of performance-influence models is measured in terms of how well
the models predict the performance of all configurations.
The main cost driver for building performance-influence models is the need to execute and measure a potentially large
number of sampled configurations -- there is typically a tradeoff between accuracy and
cost in that models trained on fewer samples are less expensive to build, but also less accurate
(e.g., if the sampled configurations do not cover all relevant execution paths, cf. Fig.~\ref{concept-cost-error}).

Essentially, all existing approaches for building performance-influence models are
black box in that they do not consider the implementation of the system.
The key differentiators are how configurations are sampled and how (and what kind of) models are fitted.
Many combinations have been explored~\citep{SRA:GPCE13, GCASW:ASE13, SGSAC:ASE15, SKKABRS:ICSE12, SRKKAS:SQJ12, SGAK:ESECFSE15}, with different tradeoffs among applicability, cost, and accuracy~\citep{KSKGA:SOSYM18}.

By contrast, we propose to analyze the system's implementation (white-box) to guide the sampling
to relevant configurations and to instrument the system to perform more fine-grained measurements, mapping
performance influence of options to specific code regions.
While there has been work on program analysis to identify how configuration options affect the execution
of a system~\citep{KMKBSBD:ESECFSE13, LKB:TSE18, AGPG:ASE15}, usually for testing or system
comprehension, beyond our own prior work with very limiting assumptions
(e.g., no data-flow interactions, and exclusive to compile-time variability)~\citep{SRA:GPCE13}, we are not aware of any
work on using program analysis to inform performance-influence modeling.
We conjecture that white-box strategies can achieve higher accuracy at lower cost, 
if the analysis identifies relevant options and interactions.

The insights on which we build this work are \emph{Irrelevance} (not all options influence the performance of a system on a given workload), \emph{Orthogonality} (not all options interact with each other), and \emph{Low Interaction Degree} (most options tend to interact only with few other options)~\citep{KMKBSBD:ESECFSE13, JSVKPA:ASE17, KBK:AOSD11, MWKTS:ASE16, SKKABRS:ICSE12, SRA:GPCE13, KSKGA:SOSYM18, LKB:TSE18, RSMFP:ICSE10, NKCFP:FSE16}.
Other insights, such as prefix sharing and variational execution~\citep{MWKTS:ASE16}, could be exploited to efficiently build accurate models.
However, these insights require special dynamic analysis techniques with excessive overhead, even for small systems, which is why we do not consider them in this work. \looseness=-1

In the following paragraphs, we describe the state of the art approaches for building performance-influence models in terms of their sampling, measuring, and learning techniques, and to what degree they exploit the insights that we consider in this work.
Table~\ref{state-of-the-art-comparison} summarizes the approaches and Table~\ref{state-of-the-art-comparison-values} compares the number of executions and accuracy of each approach when analyzing our running example (Fig.~\ref{running-example}, Lines~\ref{r:unignore1}--\ref{r:unignore2}).\looseness=-1

\begin{table}[t]
\footnotesize
\centering
\caption{Comparison of the state of the art approaches.}
\begin{threeparttable}
\begin{tabular}{llll}
\toprule
Approach & Sampling & Measuring & Learning \\ \midrule
Brute Force & Exhaustive & End-to-end & Not needed \\
SPLat & Distinct execution paths & End-to-end & Not needed \\
Sampling and Learning & Strategy based & End-to-end & Algorithm based \\
Family-Based & One configuration & Region-level & Not needed \\
{\sf ConfigCrusher} & Static analysis-based & Region-level & Not needed \\
\bottomrule
\end{tabular}
\end{threeparttable}%
\label{state-of-the-art-comparison}

\bigskip

\footnotesize
\centering
\caption{Comparison of the cost and accuracy of the state of the art approaches when analyzing the running example in Fig.~\ref{running-example}.}
\begin{threeparttable}
\begin{tabular}{lccclrl}
\toprule
 & \multicolumn{3}{c}{Insights} & & \multicolumn{2}{c}{Quality} \\ \cline{2-4} \cline{6-7}
Approach & Irrelevance & Orthogonality & LID & & Cost & Accurarcy \\ \midrule
Brute Force & \ding{55} & \ding{55} & \ding{55} & & 1024 & High \\
SPLat & \ding{51} & \ding{55} & \ding{55} & & 6 & High \\
Sampling and Learning & \ding{55} & \ding{51} & \ding{51} & & ---\tnote{1} & ---\tnote{1} \\
Family-Based & \ding{55} & \ding{51} & \ding{55} & & 1 & ---\tnote{2} \\
{\sf ConfigCrusher} & \ding{51} & \ding{51} & \ding{51} & & 4 & High \\
\bottomrule
\end{tabular}
LID = Low Interaction Degree.
\begin{tablenotes}[flushleft]
\item[1] Depends on sampling strategy, such as t-wise sampling~\citep{MKRGA:ICSE16}
\item[2] High accuracy in the absence of data-flow interactions
\end{tablenotes}
\end{threeparttable}%
\label{state-of-the-art-comparison-values}
\end{table}

\textit{Brute Force} is a black-box approach that samples \emph{all} configurations of a system and measures the execution time of the system for a given workload \emph{end-to-end}.
It is rarely used in practice due to its obvious scalability issues for all but the smallest configuration spaces.
Learning a performance-influence model from Brute Force executions is not necessary as we know the execution time of all configurations, although a simplified model could be learned for understanding~\citep{KSKGA:SOSYM18}.
In our running example, among other inefficiencies, it will execute irrelevant configurations (e.g., all configurations that explore all values of options \texttt{D} -- \texttt{J}).\looseness=-1

\textit{SPLat~\citep{KMKBSBD:ESECFSE13}}
is a white-box testing approach that we repurpose for performance analysis.\footnote{Though SPLat was designed for unit testing software product lines, the algorithm can be used to reduce the number of configurations to sample.} By instrumenting the system, it dynamically tracks the configurations that produce distinct execution paths.
It reexecutes the system until all configurations with distinct paths are explored.
While it ignores irrelevant options, since they do not produce different paths, it explores all combinations of options that it encounters during execution; each time an option is reached in a new path, it explores both values for that option.
Essentially, it is an \emph{improved version} of the Brute Force approach.
SPLat does infer from control-flow interactions that some options are only reachable when specific values are selected.
In our running example, it will explore option \texttt{C} only when option \texttt{A} is enabled.
Despite this benefit, it still often produces very large sets of configurations to sample, which can lead to scalability issues.
Similar to the Brute Force approach, it measures the execution time \emph{end-to-end} and learning a performance-influence model is not necessary. \looseness=-1

\textit{Sampling and Learning} approaches are combinations of a sampling technique; such as random sampling, feature-wise, pair-wise~\citep{MKRGA:ICSE16}, design of experiments~\citep{M:DAE06}, or combinatorial sampling~\citep{AKTLS:GPCE16,NL:CSUR11,HBG:SRE11,HMGB:IST16,HNADPB:ESE18}, to measure the \emph{end-to-end} execution time of a \emph{subset of the configuration space}, and a learning technique; such as regression, classification and regression trees~\citep{GCASW:ASE13, SGSAC:ASE15, SKKABRS:ICSE12, SRKKAS:SQJ12, SGAK:ESECFSE15}, or Gaussian Processes~\citep{JVKSK:SEAMS17}, to \emph{extrapolate} a performance-influence model. 
The number of samples and accuracy of the learned model depend on the sampling strategy and learning algorithm.
Although some sampling strategies rely on a coverage criteria to sample specific interaction degrees, such as t-wise sampling~\citep{NL:CSUR11,MKRGA:ICSE16}, they might miss important interactions leading to inaccurate models.
In addition, due to their lack of insight of the internals of the system, none of these approaches recognizes irrelevant options.\looseness=-1

\textit{Family-Based Performance Measurement~\citep{SRA:GPCE13}}, our own prior work, is currently the only white-box performance-influence modeling approach we are aware of. It uses a static mapping between options to code regions and instruments the system to \emph{measure the execution time spent in the regions}. 
Subsequently, it \emph{executes the system once} with all options enabled, tracking how much each option contributes to the execution time.
The approach works well when all options only contribute extra behavior, but do not interact.
Current implementations, however, derive the static map from compile-time variability mechanisms (preprocessor directives) \citep{SRA:GPCE13} and could not handle our running example with load-time variability (i.e., loading and processing options in variables at runtime).
Furthermore, the static map only covers direct control-flow interactions from nested preprocessor directives, and can lead to inaccurate models when data-flow interactions occur.
In our running example (Fig.~\ref{running-example}), data-flow analysis is needed to detect that the second if statement indirectly depends on option \texttt{A} (with implicit data-flow through variable \texttt{x}), leading to inaccurate performance-influence models otherwise.\looseness=-1

All the surveyed approaches build performance-influence models with different levels of applicability, cost, and accuracy, but they either overapproximate or underapproximate the interactions in a system and configurations that need to be executed to build an accurate performance-influence model.
Furthermore, none of the approaches can associate the resulting performance-influence model with regions in the source code, which can help to understand and debug individual components of a system.
The only exception is the family-based approach, but it has severe limitations and assumptions of the systems it can analyze.\looseness=-1

\subsection{ConfigCrusher}
We introduce {\sf \textbf{ConfigCrusher}}, a new white-box approach that exploits the insights of \emph{Irrelevance}, \emph{Independence}, and \emph{Low Interaction Degree}, which leads to a reduction in the cost to measure performance while also generating accurate and informative performance-influence models.
Our approach improves upon the state of the art of performance-influencing modeling by using a \emph{static analysis} to identify how \emph{load-time configuration options} may influence regions in the system through \emph{control-flow and data-flow dependencies}. Then, it derives a set of relevant configurations to \emph{measure the execution time of regions} and builds \emph{local performance-influence models} that describe how options influence the execution time of those regions.
Subsequently, the local models are aggregated to obtain a global performance-influence model for the system.

In our running example, {\sf ConfigCrusher} will identify $3$ \emph{regions affected by configuration options} (Fig.~\ref{running-example}) and use the options that influence the regions to \emph{compress} the configuration space into a set of $4$ \emph{configurations} to be sampled (Table~\ref{configuration-performance}).
Next, it will \emph{instrument} the system's regions, reducing the instrumentation overhead through additional optimization (Sec.~\ref{instrument}).
Finally, it will \emph{build} local performance-influence models (Sec.~\ref{build}) for each region based on the performance observed when executing the instrumented system with the compressed set of configurations (Table~\ref{configuration-performance}) and, subsequently, aggregate them to produce an accurate global \emph{performance-influence model}. \looseness=-1


\section{ConfigCrusher}
\label{approach}

\begin{figure}[t]
\begin{center}
\includegraphics[width=0.9\columnwidth]{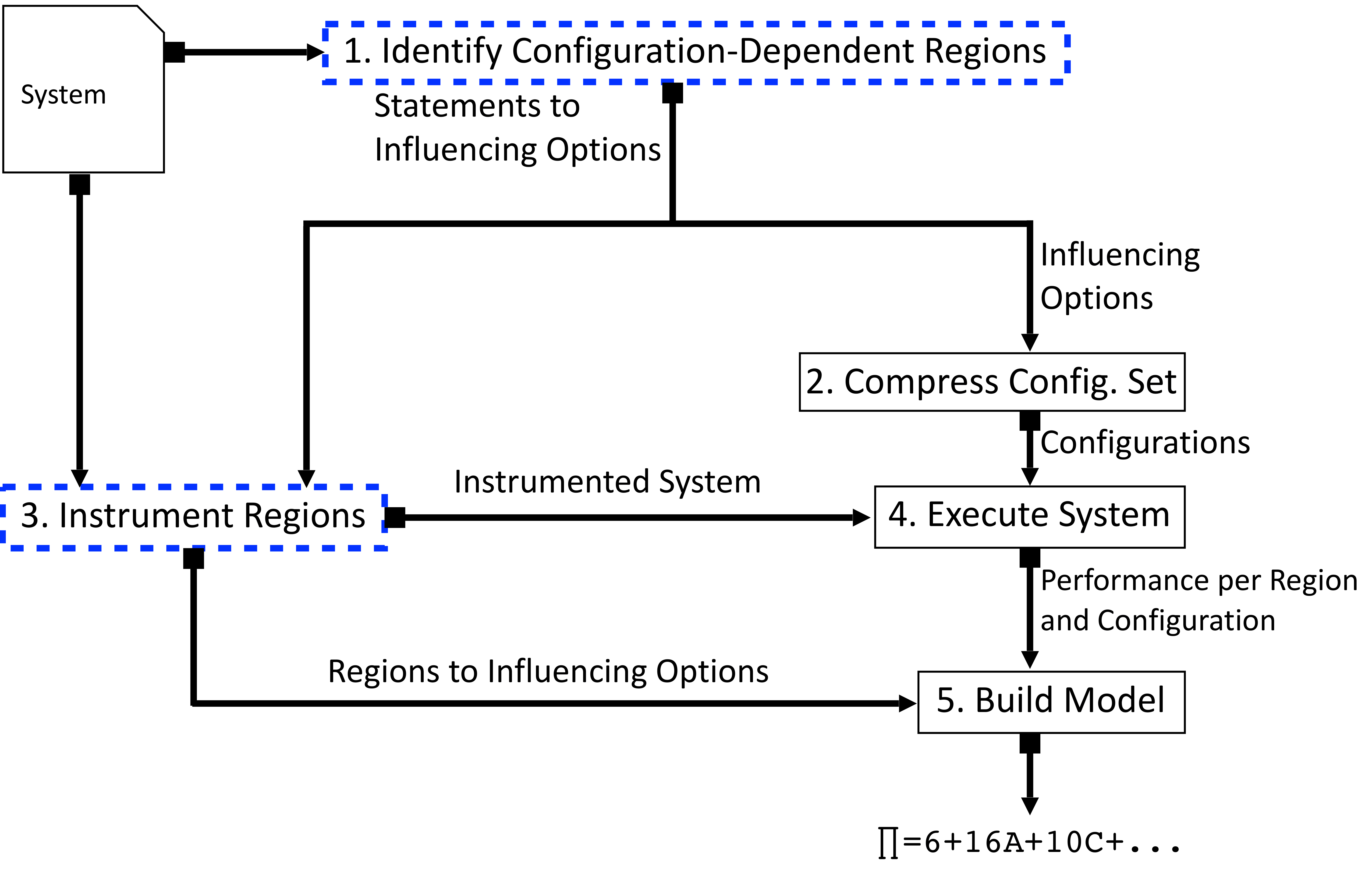}
\end{center}
\caption{Overview of {\sf ConfigCrusher}'s modular components. The components represented with solid boxes can be reused for analyzing systems implemented in any programming language. The dashed blue boxes indicate components that have to target specific programming languages.}
\label{approach-fig}
\end{figure}
\setlength{\textfloatsep}{10pt}%

The general idea of {\sf ConfigCrusher} is to identify the regions (sets of statements influenced by a set of options from control-flow and data-flow dependencies) in the system that depend on configuration options, and use these options to generate a compressed set of configurations. 
The set is then used to measure the regions' performance to build an accurate performance-influence model.
We proceed in five steps:
\begin{itemize}[noitemsep,topsep=0pt]
\item \emph{Identify Configuration-Dependent Regions} (Sec.~\ref{analysis}): We perform a data-flow analysis that identifies the control-flow statements that depend on configuration options and the code regions affected by these statements.
\item \emph{Compress Configuration Set} (Sec.~\ref{compression}): We identify the smallest set of configurations that cover all relevant executions of all regions.
 \item \emph{Instrument Regions with Optimizations} (Sec.~\ref{instrument}): We instrument the regions in the system to track their execution time in different configurations and optimize the instrumentation to reduce measurement overhead.
\item \emph{Execute the Instrumented System} (Sec.~\ref{execute}): We execute the instrumented system to measure the performance of regions.
\item \emph{Build the Performance-Influence Model} (Sec.~\ref{build}): We individually build local performance-influence models based on the measured code regions' performance and, subsequently, aggregate them to obtain a global performance-influence model.
\end{itemize}

\subsection{Identifying Configuration-Dependent Regions}
\label{analysis}

As first step, we identify the control-flow statements that depend on configuration options and the regions affected by these statements.
To this end, we create the \emph{statement influence map} $SI$ from statements~$S$ to the set of options $O$ that influence the execution of these statements $(\mathit{SI : S \rightarrow \mathcal{P}(O)})$.
We use this map later to compress a set of configurations (Sec.~\ref{compression}) and to instrument the system (Sec.~\ref{instrument}).\looseness=-1

To obtain the statement influence map, we use a data-flow analysis to track how options are used in control-flow statements.
That is, we track variables at API calls that load configuration options and then propagate them along control-flow and data-flow dependencies (including implicit flows). 
By tracking how each option flows through the system, we can identify, for each control-flow statement, the set of options that may influence this statement.
Finally, we produce the map, mapping all statements in the branches of a control-flow statement to all influencing options.\looseness=-1

\paragraph{Example:} 
The options in our running example in Fig.~\ref{running-example} (Lines~\ref{r:unignore1}--\ref{r:unignore2}) are the fields \texttt{A} -- \texttt{J}. Lines~\ref{r:noinfluence1}--\ref{r:noinfluence2} are not influenced by any options, Lines~\ref{r:influenceac1}--\ref{r:influenceac2} are influenced by the set of options $\{\texttt{A}, \texttt{C}\}$, Lines~\ref{r:influencea1}--\ref{r:influencea2} by $\{\texttt{A}\}$, and Line~\ref{r:influenceab} by $\{\texttt{A}, \texttt{B}\}$.\looseness=-1

\paragraph{}
We can reason about \emph{Irrelevance}, \emph{Orthogonality}, and \emph{Low Interaction Degree} with this data-flow analysis:
Options that influence no control-flow statements are irrelevant and never appear in the resulting map.
Likewise, we can identify which set of options interact on which control-flow statements and detect both orthogonality and low interaction degree. 
For example, in our running example, we learn that option \texttt{A} interacts with \texttt{B} and \texttt{C} separately, but not together, and that options \texttt{D} -- \texttt{J} are irrelevant in the system.\looseness=-1

\subsection{Compressing Configuration Set}
\label{compression}

Based on the statement influence map, we now calculate the \emph{compressed set of configurations} ($\mathit{CC \subseteq \mathcal{P}(O)}$) that will be executed to measure the performance of the regions in the system.
We use the set of all interactions (image of \emph{SI} from Sec.~\ref{analysis}, $IO : \mathit{\mathcal{P}(\mathcal{P}(O))}$) to generate this set of configurations and use it later to execute the instrumented system (Sec.~\ref{execute}).\looseness=-1

Intuitively, our goal is to execute the system such that \emph{each region} is executed for \emph{every combination of options} involved in that region, while minimizing the overall number of configurations to execute. Since different regions may be influenced by different options (orthogonality), we can execute them in the same configurations, in a process we call \emph{compression}. 
The challenge is similar to finding covering arrays in combinatorial interaction testing, such as covering all combinations of pairs of options~\citep{KKL:ICT13,AKTLS:GPCE16,HBG:SRE11,HMGB:IST16,HNADPB:ESE18}.
However, we need to cover different interaction strengths for different sets of options depending on which combinations of options have been detected in our statement influence map.\looseness=-1

\begin{algorithm}[t]
\footnotesize
\caption{Compression of configuration set}
\label{compression-algorithm}
	\KwIn{Influencing options $IO : \mathcal{P}(\mathcal{P}(O))$}
	\KwOut{Compressed set of configurations $CC : \mathcal{P}(C)$}
	\SetKwFunction{FCompressConfigurations}{compress\_configuration\_set($IO$)}
		
	\Fn{\FCompressConfigurations} {
		\textit{unique\_opts} := \texttt{unique\_options(}\textit{IO}\texttt{)} $\label{c:unique1}$ \tcp{Get unique sets of options} 
			
     	\tcp{Remove subsets of other sets}
	
		\textit{unique\_opts} := \texttt{remove\_subsets(}\textit{unique\_opts}\texttt{)} $\label{c:unique2}$
		
		\textit{options\_to\_confs} := \texttt{new Map()}
		
		\For{\textit{o} $\in$ \textit{unique\_opts}}{\label{c:config1}
			\textit{confs} := \texttt{configurations(}\textit{o}\texttt{)} \tcp{Get all configurations}
			
			\textit{options\_to\_conf}.\texttt{put(}\textit{o, confs}\texttt{)}
		}\label{c:config2}
		
		\textit{o}$_1$ := $\varnothing$, \textit{cs}$_1$ := $\varnothing$
		
		\For{\textit{o}$_2$, \textit{cs}$_2$ $\in$ \textit{options\_to\_confs}}{\label{c:comp1}
			\textit{CC} := $\varnothing$, \textit{pivot} := \textit{o}$_1$ $\cap$ \textit{o}$_2$

			\While{\textit{c} $\in$ \textit{cs}$_1$ $\land$ \textit{cs}$_1$.\texttt{hasNext()} $\land$ \textit{cs}$_2$.\texttt{hasNext()}}{				
				\textit{pv} := \texttt{pivot\_value(}\textit{pivot, c}\texttt{)} 				\tcp{Get value of pivot}
				
				\textit{c}$_2$ := \texttt{conf\_with\_pv(}\textit{cs}$_2$\texttt{)} 				\tcp{Get conf. with value of pivot}
								
				\textit{CC}.\texttt{add(}\textit{c}$_1$ $\cup$ \textit{c}$_2$\texttt{)}
			}

			\textit{CC}.\texttt{add\_remaining(}\textit{cs}$_1$\texttt{)}, \textit{CC}.\texttt{add\_remaining(}\textit{cs}$_2$\texttt{)}

			\textit{o}$_1$ := \textit{o}$_1$ $\cup$ \textit{o}$_2$, \textit{cs}$_1$ := \textit{CC}
		}\label{c:comp2}

		\Return{\textit{CC}}
	}\end{algorithm}

We developed a heuristic compression algorithm (Algorithm~\ref{compression-algorithm}) to find and compress a set of configurations that we use to measure the performance of the system.
First, we select all unique sets of options that are not subsets of other sets (Lines~\ref{c:unique1}--\ref{c:unique2}) and calculate all combinations of each set (Lines~\ref{c:config1}--\ref{c:config2})---these are the minimum combinations we need to cover. 
Next, we compress the set of configurations (Lines~\ref{c:comp1}--\ref{c:comp2}) by iteratively merging the partial configurations around the options that are common between two sets of options (i.e., the pivot).\looseness=-1

\paragraph{Example:} 
In our running example (Lines~$1$--$15$), the regions are influenced by the sets of options $\{A\}$, $\{A, B\}$, and $\{A, C\}$. That is, we need to cover two combinations for $\{A\}$ (with \texttt{A} enabled and disabled), four combinations of $\{A, B\}$, and four combinations of $\{A, C\}$.
The four combinations of $\{A, B\}$ already subsume the two configurations of $\{A\}$.
Furthermore, based on the pivot $\{A\}$ of the remaining sets, we can create a merged compressed set of four configurations that still cover all interactions of \texttt{A} with \texttt{B} and \texttt{A} with \texttt{C}:~$\bigl\{ \{\}, \{B, C\}, \{A\}, \{A, B, C\} \bigr\}$.\looseness=-1

\paragraph{}
Note how compression exploits  \emph{Irrelevance} and \emph{Orthogonality}: It does not consider irrelevant options (e.g., \texttt{D}) and does not consider the combinations of options that do not interact (e.g., \texttt{B} and \texttt{C}).
The size of the compressed set is dominated by the size of the largest interaction (at least $2^n$ configurations for an interaction among $n$ options; $n=2$ in our running example), which is often moderate due to \emph{Low Interaction Degree}. At the same time, independent interactions of the same size can often be merged effectively.
\looseness=-1

\subsection{Instrumenting Regions with Optimizations}
\label{instrument}

Next, we instrument the system to measure its performance broken down by code regions.
As part of the instrumentation, we identify and optimize the actual regions used for measurement, derived from the statement influence map (Sec.~\ref{analysis}).
We subsequently execute the instrumented system (Sec.~\ref{execute}) with the compressed set of configurations (Sec.~\ref{compression}) to build the performance-influence model (Sec.~\ref{build}).\looseness=-1

\begin{algorithm}[t]
\footnotesize
\caption{Identify regions}
\label{region-location-algorithm}
	\KwIn{Control-flow graph $\mathit{CFG}$, Statement-influence map $SI : S\rightarrow \mathcal{P}(O)$}
	\KwOut{System with instrumented regions $R \rightarrow \mathcal{P}(E) \times \mathcal{P}(E)$, Regions to influencing options $R \rightarrow \mathcal{P}(O)$}
	\SetKwFunction{FIdentifyRegions}{identify\_regions($CFG, SI$)}
				
	\Fn{\FIdentifyRegions} {
		\Foreach(\textit{stmt} $\in$ \texttt{statements(}\textit{CFG}\texttt{)}) {
			\textit{idom} := \texttt{idom(}\textit{stmt, CFG}\texttt{)} \label{i:rstart} 			\tcp{Get immediate dominator}
					
			\tcp{\texttt{influence(}\textit{s, SI}\texttt{)}$: S \rightarrow \mathcal{P}(O)$}
			\If{\texttt{influence(}\textit{stmt, SI}\texttt{)} $\ne$ $\varnothing$ $\wedge$ \texttt{influence(}\textit{stmt, SI}\texttt{)} $\ne$ \texttt{influence(}\textit{idom, SI}\texttt{)}} 	 {\label{i:rend1}
				\textit{r} := \texttt{new Region()}
			
				\tcp{Omit incoming edges from loops}
				\Foreach(\textit{edge} $\in$ \texttt{in(}\textit{stmt, CFG}\texttt{)}) {
					\texttt{start(}\textit{r, edge, stmt}\texttt{)} \tcp{Map $r \rightarrow e$ and $r \rightarrow \mathcal{P}(\mathcal{P}(O))$}
				}
				\textit{pdom} := \texttt{ipdom(}\textit{stmt, CFG}\texttt{)} \tcp{Get immediate post-dominator}\label{i:pdom1}
					
				\While{\texttt{influence(}\textit{stmt, SI}\texttt{)} = \texttt{influence(}\textit{pdom, SI}\texttt{)}} {
					$\mathit{pdom}$ := \texttt{ipdom(}$\mathit{pdom, CFG}$\texttt{)}
				}\label{i:pdom2}	
					
				\Foreach(\textit{edge} $\in$ \texttt{in(}\textit{pdom}\texttt{)}) {
					\texttt{end(}\textit{r, edge, stmt}\texttt{)} \tcp{Map $r \rightarrow e$ and $r \rightarrow \mathcal{P}(\mathcal{P}(O))$}
				}	
			}\label{i:rend2}	
		}
	}
\end{algorithm}

A region is a set of statements influenced by the same set of options, identified by a set of control-flow edges that start the region and another set of edges that end it.
Algorithm~\ref{region-location-algorithm} calculates the regions and their start (Line~\ref{i:rstart}) and end edges (Lines~\ref{i:rend1}--\ref{i:rend2}) in a method.
A region starts before the first statement influenced by a set of options (indicated by the statement influence map) and ends after the last statement influenced by the same set of options. 
One task of the algorithm is to find the end of a region where all the paths originating from a control-flow statement meet again (i.e., the immediate post-dominator) (Lines~\ref{i:pdom1}--\ref{i:pdom2}).
The algorithm obtains the immediate post-dominator and continuously searches for the next one until it finds the last statement with the same influence as the current control-flow statement.\looseness=-1

After identifying all regions, we instrument the start and end edges of these regions with statements to log their execution time and measure their influence on performance.
We also instrument the entry point of the system
to measure the performance of code not influenced by any options. 
The result of executing an instrumented system is the total time spent in each region.\looseness=-1

\begin{figure}[t]
\begin{flushleft}
\begin{subfigure}{0.8\columnwidth}
  \centering
  \includegraphics[width=\columnwidth]{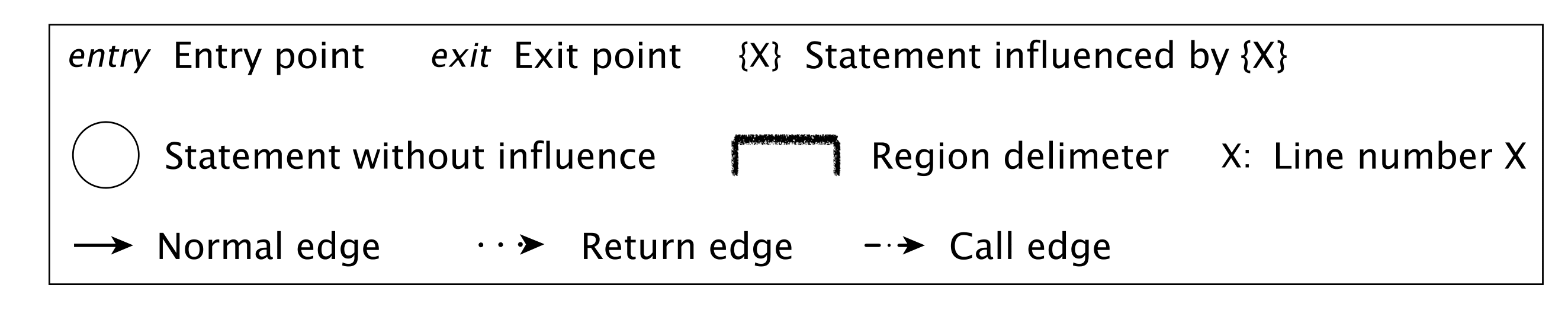}
\end{subfigure}%
\end{flushleft}
\break
\begin{subfigure}{.45\columnwidth}
  \centering
  \includegraphics[width=\linewidth]{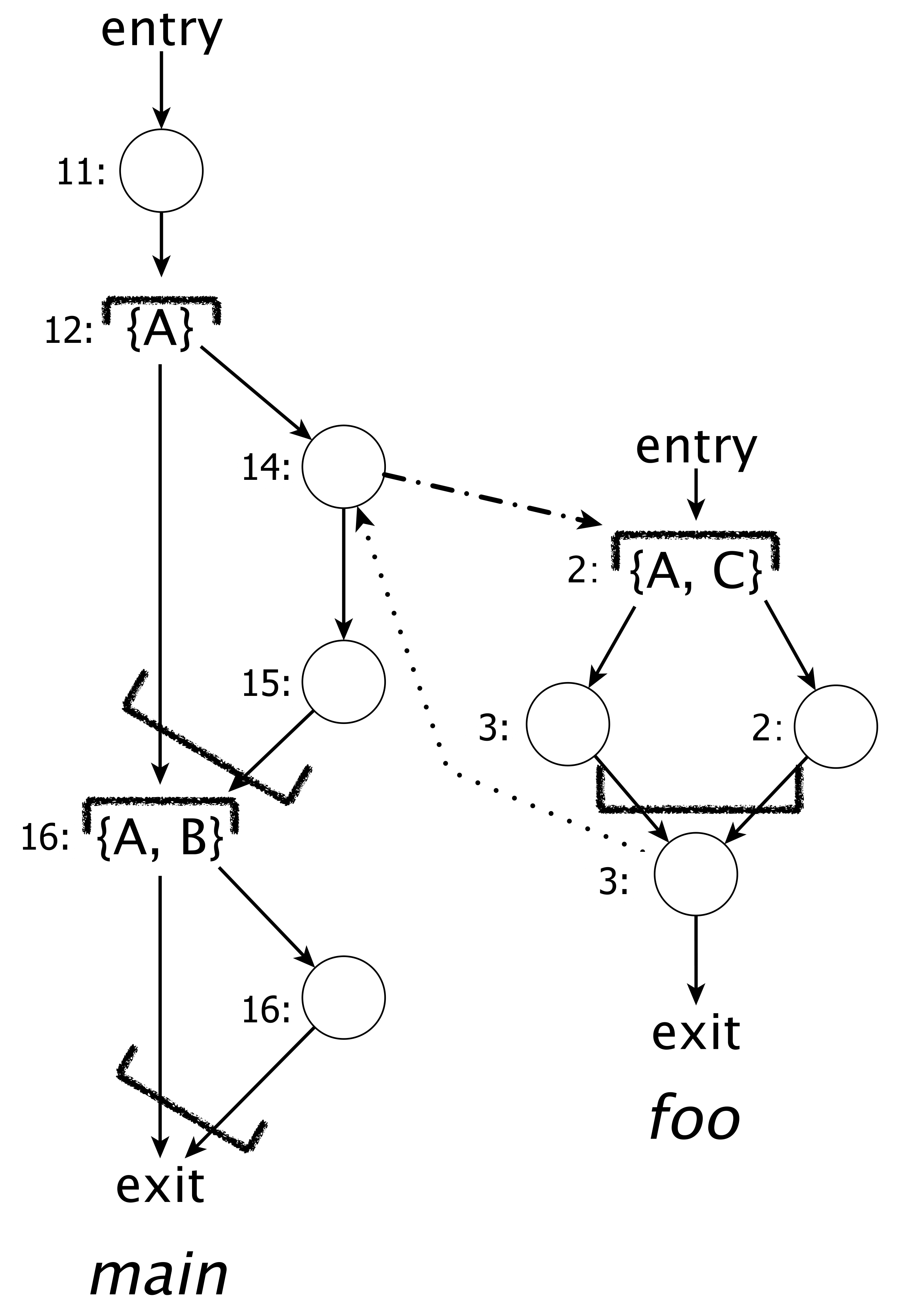}
  \caption{Unoptimized}
  \label{cg-cfg-running-example-b}
\end{subfigure}%
\hspace{10mm}
\begin{subfigure}{.45\columnwidth}
  \centering
  \includegraphics[width=\linewidth]{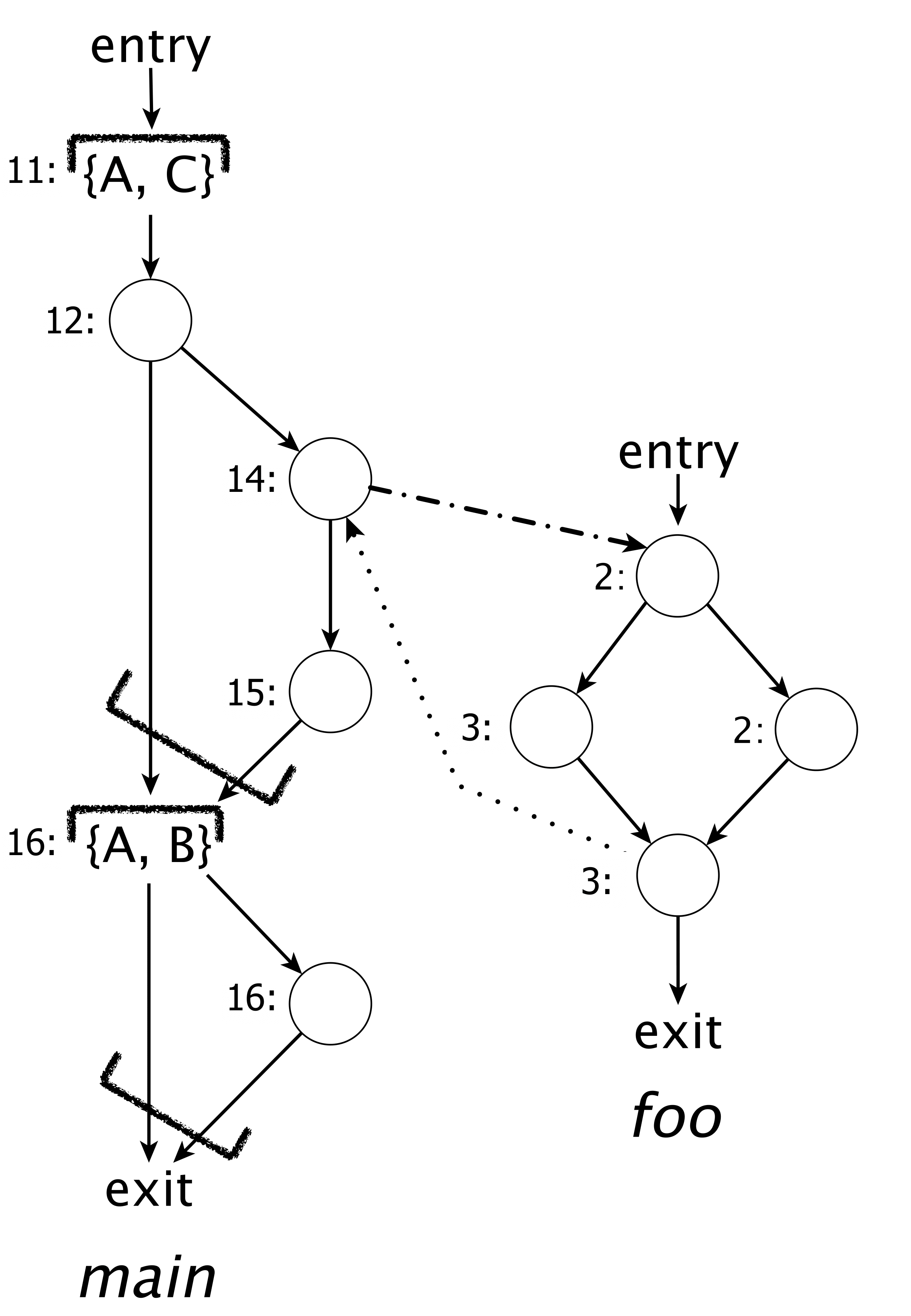}
  \caption{Optimized}
  \label{cg-cfg-running-example-c}
\end{subfigure}%
\cprotect\caption{Unoptimized and optimized instrumented control-flow graphs of the methods of Fig.~\ref{running-example}. For simplicity, statements within the regions and those before line $9$ in method \verb|main| are ignored. 
}
\label{cg-cfg-running-example}
\end{figure}

\paragraph{Example:} 
Fig.~\ref{cg-cfg-running-example}a shows the three regions that we instrument in the control-flow graphs of our running example in Fig.~\ref{running-example}.
One region contains statements $\{12, 14, 15\}$ since statement \ref{r:influencea1} is influenced by $\{A\}$ and its last post-dominator without that influence is \ref{r:influenceab}.\looseness=-1

\paragraph{Optimization.}
Although Algorithm~\ref{region-location-algorithm} can identify regions in a system, we observed excessive overhead in execution even in small systems (see Sec.~\ref{rq2-overhead}).
We found that the overhead arose from redundant, nested regions (i.e., regions with the same set of influencing options), and regions executed repeatedly in loops.
Subsequently, we identified optimizations to reduce measurement overhead through instrumenting different regions without altering the performance-influence model that we produce. 
Specifically, we perform optimizations that preserve the following two invariants.\looseness=-1

\textbf{Invariant 1 (Expand regions):} \emph{Statements not influenced by options can be added to a region without altering the performance-influence model that is generated and without increasing measurement effort}.
Statements not influenced by options contribute the same execution time to all configurations.
Therefore, including these statements in a region increases the execution time of the region equally for all configurations, but does not affect the performance difference among configurations used to build the performance-influence model.\looseness=-1

\paragraph{Example:} 
Consider the statement in Line~\ref{l:exec1} and Region $R_2$ in our running example.
The statement takes $1$ second to execute and Region $R_2$ takes $2$ seconds to execute when option \texttt{A} is enabled, from which we can derive the partial performance-influence model $\Pi_{R_2} = 1 + 2\texttt{A}$.
Since the statement is not influenced by any options, we can include it in the region, now observing $1$ or $3$ seconds executions depending on whether \texttt{A} is enabled, preserving the same $2$ seconds difference and resulting in the same model.\looseness=-1

\paragraph{}
\textbf{Invariant 2 (Merge regions):} \emph{$\mathit{GI : \mathcal{P}(\mathcal{P}(O))}$ is the set of all interactions in the system.
Two consecutive regions or an outer and an inner region influenced by interactions $\mathit{i_1 \in GI}$ and $\mathit{i_2 \in GI}$
can be merged if $\mathit{i_1 \cup i_2 \in GI}$ without altering the performance-influence model that is generated and without increasing measurement effort}.
Merging two consecutive regions or an outer and an inner region forms an interaction between the options that influence both regions.
Therefore, we have to sample all combinations of the interaction to obtain their influence on the region.
If that interaction is already present in the system, we already sample all these configurations anyway.
Therefore, we can merge these regions into one that is influenced by the interaction of the two regions.
As stated in invariant $1$, merging does not affect the absolute performance difference used to build the performance-influence model.
By merging regions, especially nested regions within loops, we \emph{significantly} reduce the number of regions that are executed, which \emph{significantly} reduces the overhead of measuring the instrumented system.\looseness=-1

\paragraph{Example:} 
Consider regions $R_2$ and $R_3$ in our running example.
Region $R_2$ is influenced by $\{A\}$ and region $R_3$ by $\{A, B\}$.
We sample all $4$ combinations of \texttt{A} and \texttt{B} to conclude that Region $R_2$ takes $2$ seconds to execute when option \texttt{A} is enabled and Region $R_3$ takes $3$ seconds when both \texttt{A} and \texttt{B} are enabled, resulting in the partial performance-influence model $\Pi_{R_{2} + R_{3}} = 2\texttt{A} + 3\texttt{AB}$.
Since we already sample all configurations for interaction $\{A, B\}$ and since $\{A\} \cup \{A, B\}$ does not create a new interaction, we can merge both regions into one that is influenced by interaction $\{A, B\}$ without having to sample more configurations.
In this case, the merged region would take $2$ seconds when \texttt{A} is enabled and $5$ seconds when both \texttt{A} and \texttt{B} are enabled, resulting in the same performance-influence model when we calculate the actual influence of enabling both \texttt{A} and \texttt{B} (i.e., $+3$ seconds).
With the same reasoning, we can also merge regions $1$ and $2$ in our running example.\looseness=-1


\begin{algorithm}[t]
\footnotesize
\caption{Propagate influence down}
\label{propagate-down-algorithm}
	\KwIn{Statement $stmt$, Control-flow graph $CFG$, Statement-influence map $SI: S \rightarrow \mathcal{P}(O)$}
	\KwOut{Optimized statement-influence map $S \rightarrow \mathcal{P}(O)'$}

	\Fn{propagate\_down($stmt, CFG, SI$)} {
		\textit{ipdom} := \texttt{ipdom(}\textit{stmt, CFG}\texttt{)} \tcp{Get immediate post-dominator}
	
		\tcp{Get set of statements in all paths}
		\textit{pstmts} := \texttt{paths\_stmts(}\textit{stmt, ipdom}\texttt{)} $-$ \textit{ipdom}

		\Foreach(\textit{ps} $\in$ \textit{pstmts}) {\label{d:rend1}
		    \tcp{\texttt{influence(}\textit{ps, SI}\texttt{)}$: S \rightarrow \mathcal{P}(O)$}
			\If{\texttt{influence(}\textit{ps, SI}\texttt{)} $\subset$ \texttt{influence(}\textit{stmt, SI}\texttt{)}} {
				\texttt{influence(}\textit{ps, SI}\texttt{)} := \texttt{influence(}\textit{stmt, SI}\texttt{)}
			}
		}\label{d:rend2}
	}
\end{algorithm}

\begin{algorithm}[t]
\footnotesize
\caption{Propagate regions up}
\label{propagate-up-algorithm}
	\KwIn{Statement $stmt$, Control-flow graph $CFG$, Statement-influence map $SI: S \rightarrow \mathcal{P}(O)$, Set of all interactions in the system $GI : \mathcal{P}(O)$}
	\KwOut{Optimized statement-influence map $S \rightarrow \mathcal{P}(O)'$}

	\Fn{propagate\_up($stmt, CFG, SI, GI$)} {
		\Foreach(\textit{pred} $\in$ \texttt{preds(}\textit{stmt, CFG}\texttt{)}) {
		    \tcp{\texttt{influence(}\textit{s, SI}\texttt{)}$: S \rightarrow \mathcal{P}(O)$}
			\If{\texttt{influence(}\textit{pred}\texttt{)} $\cup$ \texttt{influence(}\textit{stmt}\texttt{)} $\in$ \textit{GI} $\land$ \texttt{influence(}\textit{pred}\texttt{)} $\neq$ \texttt{influence(}\textit{pred}\texttt{)} $\cup$ \texttt{influence(}\textit{stmt}\texttt{)}} {
				\texttt{influence(}\textit{pred}\texttt{)} := \texttt{influence(}\textit{stmt}\texttt{)}
			}
		}
	}	
\end{algorithm}

\paragraph{}
We developed two algorithms (Algorithm~\ref{propagate-down-algorithm} and Algorithm~\ref{propagate-up-algorithm}) that use the invariants to propagate the options that influence statements up and down a control-flow graph (i.e., intraprocedually), as well as across graphs (i.e., interprocedually), to expand, merge, and pull out regions.
The propagation in Algorithm~\ref{propagate-down-algorithm} merges consecutive regions (Lines~\ref{d:rend1}--\ref{d:rend2}) and expands where regions end. 
The propagation in Algorithm~\ref{propagate-up-algorithm} pulls out nested regions and expands where regions start.
Obeying our invariants, both algorithms never create new interactions nor do they alter the performance-influence model that we generate, but significantly reduce the overhead of measuring the instrumented system.
After propagation, we identify the regions and instrument them as before (Algorithm~\ref{region-location-algorithm}).\looseness=-1

The propagation algorithms are non-deterministic (i.e., different results are obtained depending on the order in which regions are merged).
In fact, different orderings can be used to optimize for different goals.
Assuming that most of the overhead occurs in nested regions, especially those inside loops, we prioritize pulling regions out of loops. 
(We experimented with other orderings and the results were similar to Table~\ref{region-reduce}).
Fig.~\ref{cg-cfg-running-example}b presents an optimized instrumentation that prioritizes our goal, in which we pulled out the region in the callee.\looseness=-1

\subsection{Executing the Instrumented System}
\label{execute}

After instrumentation, we can now execute the system with the compressed set of configurations ($CC$, Sec.~\ref{compression}) and track execution times for each region. We produce a \emph{configuration performance map} $CP$, which maps each region $R$ in each executed configuration to a corresponding execution time $T$ ($\mathit{CP : CC \rightarrow (R \rightarrow T)}$).\looseness=-1

At the start and end of every region, we record the current time and log the difference as the execution time of the region. 
Since regions might be nested during execution, we also keep a stack of regions at runtime and subtract the time of nested regions from the time of outer regions.
This additional step can become a source of overhead for deeply nested regions, which is what we observed in the unoptimized instrumented systems.
We tried building a trace of regions and processing the execution times after the system finished executing.
However, due to the large number of regions that were executed, the systems ran out of memory.
Our evaluation shows that the dynamic processing incurs low overhead (Sec.~\ref{rq2-overhead}).\looseness=-1

\begin{table}[t]
\footnotesize
\centering
\caption{Configuration performance map of the optimized region of Fig.~\ref{running-example}. For simplicity, the measurement noise was removed.}
\begin{threeparttable}
\begin{tabular}{lllllrrr}
\toprule
\multicolumn{4}{c}{Configurations} & \multicolumn{1}{c}{} & \multicolumn{3}{c}{Regions} \\ \cline{1-4} \cline{6-8} 
A & B & C & D &  & Base (s) & OR$_{1}$ $\equiv$ \{A, C\} (s) & OR$_{2}$ $\equiv$ \{A, B\} (s) \\ \midrule
F & F & F & F &  & 1 & 0 & 0 \\
F & T & T & F &  & 1 & 0 & 0 \\
T & F & F & F &  & 1 & 3 & 0 \\
T & T & T & F &  & 1 & 6 & 3 \\
\bottomrule
\end{tabular}
OR$_{1}$: Optimized Region $1$;
OR$_{2}$: Optimized Region $2$.
\end{threeparttable}%
\label{configuration-performance}
\end{table}

\paragraph{Example:} 
Table~\ref{configuration-performance} presents a configuration performance map for the two optimized regions and compressed set of four configurations of our running example.\looseness=-1

\subsection{Building the Performance-Influence Model}
\label{build}

Our final step is to build the \emph{performance-influence model} $\Pi$ that predicts the performance of each configuration $(\Pi : \mathit{C \rightarrow T})$, based on the configuration performance map ($\mathit{CP : CC \rightarrow (R \rightarrow T)}$, Sec.~\ref{execute}) and the region influence map ($\mathit{RI: R \rightarrow \mathcal{P}(O)}$, Sec.~\ref{instrument}).\looseness=-1

To build the global performance-influence model, we first build local models for each region separately and subsequently aggregate them.
A local model contains performance terms for all combinations of options that are associated with the region (using $RI$), in the form
$\Pi_{r} = t_1XY + t_2X\neg Y+ t_3\neg XY + t_4\neg X\neg Y$ for a region $r$ with options \texttt{X} and \texttt{Y} (or $\Pi_r = t_4 + (t_2-t_4)X + (t_3-t_4)Y + (t_1-t_2-t_3+t_4)XY$ to highlight the influence of options and avoid negated terms)~\citep{SKKABRS:ICSE12}.
If a region has been executed multiple times for the same combination of options in one configuration, the execution time $t_i$ should not differ beyond usual measurement noise (since other options should not influence the region), thus we average the execution time. \looseness=-1

The global performance-influence model $\Pi$ is obtained by aggregating all local performance-influence models. Note that local models can be useful for understanding and debugging the individual regions in the system.\looseness=-1

\paragraph{Example:} 
With the measurement times in Table~\ref{configuration-performance}, we build the local models of the base region as $\Pi_{\texttt{Base}}$ = \verb|1| (averaged over 4 executions) and of the other two regions as $\Pi_{\texttt{OR}_{1}}$ = \verb|3A|$\neg$\verb|C + 6AC| = \verb|3A + 3AC| and $\Pi_{\texttt{OR}_{2}}$ = \verb|3AB|, resulting in the overall model $\Pi = 1 + 3\texttt{A} + 3\texttt{AB} + 3\texttt{AC}$.

\section{Implementation}
\label{implementation}

To show the feasibility of our white-box approach, we implemented {\sf ConfigCrusher} for Java systems and made it publicly available~\citep{VJSSAK:ASEJ20SM}.
Its modular design allows {\sf ConfigCrusher} to analyze systems in any programming language; only the data-flow analysis (Sec.~\ref{analysis}) and instrumentation (Sec.~\ref{instrument}) components have to target the specific language (Fig.~\ref{approach-fig}).\looseness=-1

There are several strategies to track data-flow in a system: manual tracking, static analysis~\citep{RK:ICSE11, ARFBBKLOM:PLDI14, LKB:TSE18, EGCCJMS:OSDI10, QWR:ISSTA18, DALC:ISSRE16}, and dynamic analysis~\citep{BK:SIGPLANNOT14, MWKTS:ASE16, NKCFP:FSE16, RSMFP:ICSE10, YHASFC:PLDI16, AF:POPL12, AF:PLAS09}. We used the state of the art~\citep{DALBSM:ISSTA17, QWR:ISSTA18, WZR:SOAP16, PBW:FSE18} object-, field-, context-, and flow-sensitive static taint analysis engine FlowDroid~\citep{ARFBBKLOM:PLDI14}.
A taint analysis, typically used in the security domain, tracks what variables have been affected by selected inputs (sources) and are used in specific locations (sinks). 
We annotated the API calls to load configurations options as sources and control-flow statements as sinks.\looseness=-1

We used the ASM library~\citep{BLC:AECS02} to add bytecode instructions to measure the execution time of  regions.
We used the Soot framework~\citep{VCGHLS:CASCON99} to build the call graph of a system to optimize the instrumentation of regions.

\paragraph{Limitations.}
Our {\sf ConfigCrusher} implementation is limited to analyzing single-threaded systems, as we dynamically process the execution time of regions, keeping a stack of regions at runtime to subtract the time of nested regions from the time of outer regions.
Having one region executing at a time facilitates the processing and calculations that we perform.
Other measurement strategies could be used, such as using a performance profiler, to measure the execution time of regions in multi-threaded systems.

At its core, the used static taint analysis is unsound and can lead to overtainting~\citep{QWR:ISSTA18, WZR:SOAP16}, which can affect the results of our approach.
For instance, at the one extreme, if the analysis \emph{misses all interactions}, we will fail to produce a performance-influence model.
At the other extreme, if the analysis indicates that \emph{all options interact}, our approach results in the Brute Force approach.
In addition, despite the high precision of FlowDroid~\citep{ARFBBKLOM:PLDI14}, the analysis is challenged by the size of the call graph, which restricts the size of the systems that our implementation of {\sf ConfigCrusher} can analyze~\citep{AKKGZARB:ICSE15, ARFBBKLOM:PLDI14, B:ICSENIER18, LSBM:ASE15, DALBSM:ISSTA17, QWR:ISSTA18, WZR:SOAP16, PBW:FSE18}.
Similar to other approaches~\citep{AKKGZARB:ICSE15, QWR:ISSTA18, LKB:TSE18}, we reduced the precision of some FlowDroid specific settings (e.g., used an unexceptional control-flow graph) for a faster analysis (the analysis ran out of memory for systems $3$ -- $10$ in Table~\ref{systems} using the default settings).
Despite running the static analysis on a server with $512$ GB of RAM and $32$ CPU cores,
we were forced to exclude some systems from our evaluation since the server either used all of its memory or did not finish the analysis after $4$ hours. 
Avdiienko et al.~\citep{AKKGZARB:ICSE15} experienced similar results on a server with more RAM and CPU cores.
Nevertheless, our evaluation demonstrates the feasibility of our implementation to produce accurate and informative performance-influence models, signifying that our approach is robust despite the levels of unsoundness and overtainting of FlowDroid. 
We expect that incorporating advancements in scaling taint analyses~\citep{DALBSM:ISSTA17,  B:ICSENIER18, LSBM:ASE15, CB:ASE16, BJMVDDE:ASE15, AMN:SOAP17, SAB:ACMPL17, GZL:ESECFSE17, ZS:POPL17}, the results in this article and the benefits of our approach will generalize to larger systems.
We conjecture that similarly accurate results can be achieved with other taint analysis implementations.\looseness=-1

\section{Evaluation}
\label{evaluation}
To demonstrate the feasibility and potential of our white-box approach, we evaluate {\sf ConfigCrusher} against state of the art approaches to build performance-influence models, for a specific workload, input size, and underlying hardware, in terms of the cost (i.e., number of configurations to sample and time to execute those configurations) to generate performance-influence models and their accuracy.
Subsequently, we explore the usefulness of {\sf ConfigCrusher}'s local performance-influence models to identify the local influence of options on performance.
Specifically, we address the following research questions:\looseness=-1

\vspace{1ex}

\textit{\textbf{RQ1: How does {\sf \textbf{\textit{ConfigCrusher}}} compare to other performance-influence modeling approaches in terms of cost and accuracy?}} 
We compare the effectiveness of {\sf ConfigCrusher} regarding the cost of generating models and their accuracy to state of the art black-box and white-box approaches. \looseness=-1

\vspace{1ex}

\textit{\textbf{RQ2: How much overhead is induced by instrumentation?}}
One of the goals of {\sf ConfigCrusher} is to build performance-influence models efficiently.
As discussed in Sec.~\ref{instrument}, we observed an excessive amount of overhead when executing our unoptimized instrumented systems.
We evaluate the effectiveness of our optimization by exploring how much overhead the instrumented regions induce and how it affects the performance that we measure.\looseness=-1

\vspace{1ex}

\textit{\textbf{RQ3: To how many regions can the influence of options on performance be localized?}}
One of the benefits of {\sf ConfigCrusher} over state of the art approaches is that it builds local performance-influence models, which indicate whether and how the options locally influence the performance of a system.
In an exploratory analysis, we examine the local performance-influence models to determine the local influence of options on performance.
Subsequently, we analyze the source code regions corresponding to the local models to further investigate how they are influenced by options.
We conjecture that this type of information, derived from the local models, can provide insights to developers and maintainers for enhanced analysis of individual components of a system.\looseness=-1

\subsection{Subject Systems}
\label{subject-systems}
The subject systems are summarized in Table~\ref{systems}.
We selected a representative set of configurable systems that satisfy the following criteria: 
(a)~systems from a variety of domains to increase external validity, (b)~systems with at least $5$ options (the Brute Force approach would produce results cheaply for systems with few options), (c)~systems with characteristics representative of large-scale configurable systems (e.g., systems with binary and non-binary options),
(d)~single-threaded systems with deterministic execution time (we sampled each system multiple times with different approaches and observed execution times within usual measurement noise), and (e)~systems for which the static taint analysis terminated. 
We included systems that have been used in previous studies, for comparability of results (systems $4$, $7$, $8$)~\citep{KMKBSBD:ESECFSE13, SRA:GPCE13, SDG:ICSE17}, and new configurable systems with a total of $860$+ stars and $155$+ forks on GitHub, and used by $160$+ open-source projects, at the time of writing, to showcase the applicability of our approach ($2$, $3$, $5$, $6$, $8$, $9$, $10$).
In the following sections, we consider the entire system of Fig.~\ref{running-example} (Lines~\ref{p:whole1}--\ref{p:whole2}) as the Running Example~(system $1$ in Table~\ref{systems}) to showcase the potential of our approach.\looseness=-1

\begin{table}[t]
\footnotesize
\centering
\caption{Configurable subject systems.}
\begin{threeparttable}
\begin{tabular}{rllrrrr}
\toprule
ID & Name & Domain & \# SLOC & \# Opt. & \# Conf. & CID \\ \midrule
1 & Running Example & Example & 69 & 10 & 1 024 & --- \\
2 & Pngtastic Counter & Processing & 1 250 & 5 & 32 & \texttt{7f96382}\\
3 & Pngtastic Optimizer & Optimization & 2 553 & 5 & 32 & \texttt{7f96382} \\
4 & Elevator & SPL--Benchmark & 575 & 6 & 20 & --- \\
5 & Grep\tnote{1} & Command line & 2152 & 7 & 128 & \texttt{ef9eaa7} \\
6 & Kanzi & Compression & 20 537 & 7 & 128 & \texttt{4dae29e} \\
7 & Email & SPL--Benchmark & 696 & 9 & 40 & --- \\
8 & Prevayler & Database & 1 328 & 9 & 512 & \texttt{5be1ca4} \\
9 & Sort\tnote{1} & Command line & 2 163 & 12 & 4 096 & \texttt{ef9eaa7} \\
10 & Density Converter & Processing & 1 359 & 22 & 2$^{22}$ & \texttt{2ba8373} \\
\bottomrule
\end{tabular}
CID = Commit ID.
\begin{tablenotes}
\item[1] Java implementation of the Unix command.
\end{tablenotes}
\end{threeparttable}%
\label{systems}
\end{table}

Due to their novelty, white-box approaches impose strict limitations on the systems they can analyze~\citep{SRA:GPCE13, KMKBSBD:ESECFSE13, SDG:ICSE17}.
{\sf ConfigCrusher} lifts some of these limitations; we consider data-flow interactions and do not limit the analysis to specific system implementations, which expands the types of systems that can be analyzed and increases the accuracy of the results.
Still, the used implementation of static code analysis imposes limitations on the size of systems and their number of configuration options.
We acknowledge the size of the real-world systems used in the evaluation and that black-box approaches can analyze larger systems.
However, at this stage, we want to showcase the feasibility, benefits, and potential of white-box analyses and expect that, with improvements to the used data-flow analysis~\citep{DALBSM:ISSTA17,  B:ICSENIER18, LSBM:ASE15, CB:ASE16, BJMVDDE:ASE15, AMN:SOAP17, SAB:ACMPL17, GZL:ESECFSE17, ZS:POPL17}, our implementation will analyze larger systems.
Nevertheless, we selected systems for our evaluation with representative characteristics of larger configurable systems (i.e., we observed the insights of \emph{Irrelevance, Orthogonality, and Low-Interaction Degree} of configurable systems). 
Hence, the systems that we selected are suitable to answer our research questions and we conjecture that we can obtain similar results (Sec.~\ref{rq1-comparison}) in larger systems with a more scalable implementation of the static taint analysis.
Note the general trend in the results (Sec. \ref{rq1-comparison}): all other state of the art white-box approaches have the same scalability problem.
Still, {\sf ConfigCrusher} was able to analyze real-world systems which the other approaches could not; SPLat did not scale to Density Converter~(system $10$ in Table~\ref{systems}) and the family-based approach could not analyze any system besides the software product lines.\looseness=-1 

Although our static analysis correctly identified that all options interact in Elevator~($4$), since it was purposely built with such behavior~\citep{MWKTS:ASE16, KMKBSBD:ESECFSE13, SDG:ICSE17} (i.e., our approach equals the Brute Force approach), we included the system in the evaluation since it is one of the two systems that the family-based approach can analyze.

\subsection{RQ1: Comparison to State of the Art}
\label{rq1-comparison}
With RQ1, we evaluate the cost and accuracy of the performance-influence models generated by {\sf ConfigCrusher} and how it compares to state of the art black-box and white-box approaches.
To answer this question, we measured the cost and prediction error of {\sf ConfigCrusher} and all other approaches and compared them to the ground truth. \looseness=-1

\paragraph{Procedure:} 
We established ground truth by measuring the performance of the entire configuration space four times and averaged the performance of each configuration.
Due to the high number of configurations and execution time of Sort~($9$) and Density Converter~($10$), we randomly sampled a large number of configurations each to act as the ground truth.
We observed no variation in the errors of the results presented in Table~\ref{cost-compare} and Table~\ref{error-compare} when using more than $1000$ configurations.\looseness=-1
 
Specifically, we compared {\sf ConfigCrusher} to feature-wise sampling (i.e., enable one option at a time) and pair-wise sampling (i.e., cover all combinations of all pairs of options)~\citep{MKRGA:ICSE16} with stepwise linear regression~\citep{SGSAC:ASE15, SKKABRS:ICSE12, SRKKAS:SQJ12, SGAK:ESECFSE15}, Brute Force, SPLat~\citep{KMKBSBD:ESECFSE13}, and the family-based approach~\citep{SRA:GPCE13}.
We excluded random sampling since research~\citep{JVKSK:SEAMS17, JSVKPA:ASE17, MKRGA:ICSE16, SGAK:ESECFSE15} has shown that it requires numerous samples to make accurate predictions and it is not clear how many configurations to sample for a specific system.\looseness=-1

We conjectured that SPLat behaves essentially like the Brute Force approach in all but software product lines, since all configuration options are read at the start of the system.
We included a SPLat variant, called SPLatDelayed (SAD), for which we modified the source code of the systems to delay the evaluation of options in control-flow statements~\citep{S:LJ17}.
The source code refactoring allowed us to evaluate how SPLat would operate in systems 
if it could detect when options are actually evaluated in control-flow decisions.\looseness=-1

The static taint analysis and the performance measurements were executed on a $2.2$ GHz Intel Core i$7$ MacBook Pro with $16$ GB of RAM running OS X $10.13$ (i.e., fixed underlying hardware). 
For each configuration, we initiated one VM invocation and ran the configuration~\citep{GBE:SIGPLANNOT07}.
We used the JVM options \texttt{"}\texttt{-Xms10G} \texttt{-Xmx10G} \texttt{--XX:+UseConcMarkSweepGC}\texttt{"} to reduce the overhead of garbage collection. To control for measurement noise, we measured each configuration five times and averaged the performance of each configuration.
For each system, we extracted the configuration options from the projects' documentation and executed a representative test scenario and workload provided by the system (i.e., fixed workload and input size)~\citep{VJSSAK:ASEJ20SM}.
Following the evaluation of state of the art approaches \citep{LKB:TSE18, MWKTS:ASE16, SRA:GPCE13, GCASW:ASE13, SGSAC:ASE15, SKKABRS:ICSE12, SRKKAS:SQJ12, SGAK:ESECFSE15, MKRGA:ICSE16, AKTLS:GPCE16, HNADPB:ESE18, KMKBSBD:ESECFSE13}, we discretized the non-binary options (e.g., pick either the lowest or highest value) to reduce the number of samples to execute.\looseness=-1

\paragraph{Cost Metric.} 
We measured the number of configurations and sampling time to generate a model. For {\sf ConfigCrusher}, we also measured the one-time overhead of the static analysis.\looseness=-1

\paragraph{Error Metric.} 
We used the Mean Absolute Percentage Error (MAPE) to measure the mean difference between the values predicted by a model and the values actually observed (i.e., ground truth).
For each approach, we calculate the prediction error on the configurations that the approach did not sample.
We also calculated the error across all configurations~\citep{VJSSAK:ASEJ20SM}.
We used the multiple comparison \textbf{\~T}-procedure~\citep{KHB:EJS12} with $95\%$ confidence to compare statistical differences between {\sf ConfigCrusher}'s prediction error to the prediction error of each of the other approaches.\looseness=-1

\begin{table}[t]
\footnotesize
\centering
\caption{Cost of building performance-influence models.}
\begin{threeparttable}
\begin{tabular}{rrrrrrr}
\toprule
S & BF/SA & FW & PW & SAD & FB\tnote{1} & CC\tnote{2} \\ \midrule
1 & 1024 [1.9h] & \cellcolor{blue!25} 10 [16.6s] & 56 [2.2m] & 512 [56.9m] & N/A & \cellcolor{blue!25} 8 [33.8s, 4.4s] \\
2 & 32 [2.9m] & \cellcolor{blue!25} 5 [27.2s] & 16 [1.5m] & 24 [2.2m] & N/A & \cellcolor{blue!25} 4 [21.9s, 7.8s] \\
3 & 32 [42.2m] & \cellcolor{blue!25} 5 [1.6m] & 16 [10.0m] & 16 [21.0m] & N/A & 10 [10.7m, 30.6s] \\
4 & 20 [10.8m] & \cellcolor{blue!25} 3 [50.0s] & 9 [3.3m] & 20 [10.8m] & \cellcolor{blue!25} 1 [49.5s] & 64 [---] \\
5 & 128 [10.6m] & \cellcolor{blue!25} 7 [22.1s] & 29 [1.9m] & 48 [3.5m] & N/A & 64 [5.1m, 10.2s] \\
6 & 128 [1.2h] & \cellcolor{blue!25} 7 [1.5m] & 29 [8.8m] & 64 [35.4m] & N/A & 64 [35.4m, 12.6s] \\
7 & 40 [16.9m] & \cellcolor{blue!25} 4 [23.5s] & 11 [1.7m] & 40 [16.9m] & \cellcolor{blue!25} 1 [1.1m] & \cellcolor{blue!25} 8 [1.5m, 12.8s] \\
8 & 512 [3.7h] & \cellcolor{blue!25} 9 [2.7m] & 46 [16.0m] & 144 [1.5h] & N/A & 32 [14.5m, 12.6s] \\
9 & 1298 [18.4h] & \cellcolor{blue!25} 12 [13.1m] & 79 [1.4h] & 48 [42.8m] & N/A & 256 [3.7h, 21.6s] \\
10 & 1414 [14.7h] & \cellcolor{blue!25} 22 [21.3m] & 254 [4.1h] & +24h\tnote{3} & N/A & 256 [2.1h, 42.1s] \\
\bottomrule
\end{tabular}
S = Subject system;
FW = Feature-wise;
PW = Pair-wise;
BF = Brute Force;
SA = SPLat;
SAD = SPLat Delayed;
FB = Family-Based;
CC = {\sf ConfigCrusher}.\\
A \colorbox{blue!25}{cell} indicates approaches with the lowest costs.
\begin{tablenotes}[flushleft]
\item[1] Not applicable to systems without static map derived from compile-time variability.
\item[2] Time includes the overhead of the static taint analysis.
\item[3] No data was collected due to timeout.
\end{tablenotes}
\end{threeparttable}%
\label{cost-compare}

\bigskip
\footnotesize
\centering
\caption{MAPE comparison.}
\begin{threeparttable}
\begin{tabular}{rrrrrr}
\toprule
S & Feature-wise & Pair-wise & SPLat Delayed & Family-Based\tnote{1} & {\sf ConfigCrusher} \\ \midrule
1 & 56.9$\uparrow$ & 6.2$\uparrow$ & 0.2$\uparrow$ & N/A & \cellcolor{green!40} 0.1 \\
2 & \cellcolor{green!40} 0.8\phantom{$\uparrow$} & 2.0$\uparrow$ & \cellcolor{green!40} 1.3\phantom{$\uparrow$} & N/A & \cellcolor{green!40} 1.1 \\
3 & 19.7$\uparrow$ & \cellcolor{green!40} 0.9\phantom{$\uparrow$} & \cellcolor{green!40} 1.0\phantom{$\uparrow$} & N/A & \cellcolor{green!40} 1.1 \\
4 & 51.1\phantom{$\uparrow$} & \cellcolor{green!40} 1.5\phantom{$\uparrow$} & $\varnothing$ & \cellcolor{green!40} 2.7 & $\varnothing$ \\
5 & 32.1$\uparrow$ & 114.7$\uparrow$ & \cellcolor{green!40} 1.9$\downarrow$ & N/A & \cellcolor{green!40} 3.6 \\
6 & \cellcolor{green!40} 1.9\phantom{$\uparrow$} & \cellcolor{green!40} 1.3\phantom{$\uparrow$} & \cellcolor{green!40} 1.21\phantom{$\uparrow$} & N/A & \cellcolor{green!40} 2.7 \\
7 & 100$\uparrow$ & 44.2$\uparrow$ & $\varnothing$ & \cellcolor{green!40} 2.3$\downarrow$ & 23.0 \\
8 & 111.2$\uparrow$ & 29.2$\uparrow$ & \cellcolor{green!40} 3.0$\downarrow$ & N/A & \cellcolor{green!40} 9.2 \\
9 & 90.0$\uparrow$ & 653.0$\uparrow$ & 2.4$\uparrow$ & N/A & \cellcolor{green!40} 1.6 \\
10 & 635.2$\uparrow$ & 218.9$\uparrow$ & N/A\tnote{3} & N/A & \cellcolor{green!40} 4.3 \\
\bottomrule
\end{tabular}
A \colorbox{green!40}{cell} indicates approaches with statistically indistinguishable lowest errors. $\uparrow$ approach with statistically~$>$ error than {\sf ConfigCrusher}. $\downarrow$ approach with statistically~$<$ error than {\sf ConfigCrusher}. $\varnothing$ approach sampled all configurations, thus no performance to predict.
\begin{tablenotes}[flushleft]
\item[1] Not applicable to systems without static map derived from compile-time variability.
\end{tablenotes}
\end{threeparttable}%
\label{error-compare}
\end{table}

\paragraph{Results:} 
We show the cost results in Table~\ref{cost-compare} and the error results in Table~\ref{error-compare}.
{\sf ConfigCrusher}'s prediction error is statistically indistinguishable or lower than other approaches. Furthermore, {\sf ConfigCrusher}'s high accuracy is usually achieved with lower cost compared to the other accurate approaches. Our results support our conjecture on the cost and prediction error comparison of Fig.~\ref{concept-cost-error}.\looseness=-1

Though feature-wise and pair-wise sampling tended to have lower costs than {\sf ConfigCrusher}, when their errors are taken into account, we can conclude that more configurations had to be sampled to make accurate predictions.
By comparison, for those systems, {\sf ConfigCrusher} sampled more configurations, but attained significantly lower errors.\looseness=-1

As we conjectured, SPLat behaves essentially like the Brute Force approach in all but software product lines.
We also conjectured that SPLatDelayed would produce the lowest error since it uses a heuristic to perform a more efficient Brute Force approach.
For the Running Example~($1$), Pngtastic Counter~($2$), Pngtastic Optimizer~($3$), and Sort~($9$), in which other approaches besides SPLatDelayed produced lower errors, but statistically indistinguishable, we can attribute the results to measurement noise.
Interestingly, SPLatDelayed did not finish analyzing Density Converter~($10$) within $24$ hours.
In this case, most options are read sequentially (similar to reading all options at the beginning of the system), thus indicating the limitations of the approach.\looseness=-1

Only for Elevator~($4$) and Email~($7$),  
the family-based approach remains the most efficient and accurate approach, but, at the same time, the most limited one in terms of applicability.

\paragraph{Characteristics of interactions discussion:} 
Thanks to {\sf ConfigCrusher}, we observed and confirmed the insights of \emph{Irrelevance}, \emph{Orthogonality}, and \emph{Low Interaction Degree} of configurable systems. 
In $9$ out of $10$ systems, it significantly reduced the configuration space to sample, thus reducing the cost to build accurate models.
For example, our analysis identified the irrelevant options in our running example~($1$), Grep~($5$), and Sort~($9$), which is not leveraged by the black-box approaches before sampling.
Similarly, {\sf ConfigCrusher} identified orthogonal interactions and leveraged low interaction degree to sample fewer configurations, which was not exploited by the white-box approaches.\looseness=-1

For Pngtastic Counter~($2$), Pngtastic Optimizer~($3$), and Kanzi~($6$), the black-box approaches produced accurate models with low cost.
Upon inspection of the results, we discovered that (a)~in Pngtastic Counter, the options did not affect the performance of the system; the execution time was essentially the same for all configurations, and (b)~in Pngtastic Optimizer and Kanzi, the options did affect the performance, but the execution times were clustered in a few groups.
For example, the performance of Kanzi under all configurations was either ${\sim}4$ seconds or ${\sim}61$ seconds.
We consider these three systems as outliers since previous empirical studies~\citep{SGAK:ESECFSE15, ABKS:FOSPL13, JVKSK:SEAMS17, KSKGA:SOSYM18} have shown that the performance of most configurable systems changes based on the selected configurations.\looseness=-1

\paragraph{Source of prediction error discussion:} 
Regarding {\sf ConfigCrusher}'s prediction error of Email~($7$), the system has a feature model~\citep{ABKS:FOSPL13} that describes its valid configurations.
Since the invalid configurations were not executed, {\sf ConfigCrusher} did not have all the information for each region to generate an accurate model.
Despite missing information, {\sf ConfigCrusher} was able to produce more accurate results than the other approaches, except the family-based approach.
We hope to incorporate information from a feature model to produce more accurate models for this type of systems.\looseness=-1

Regarding {\sf ConfigCrusher}'s prediction error of Prevayler~($8$), we observed that the execution time of certain regions differed beyond usual measurement noise.
This behavior occurs when the correct interaction in the regions was not captured by the static analysis (a possible consequence of the unsoundness of the used taint analysis).
We were unable to manually determine the correct interaction of the problematic regions.
We conjecture that, since the system writes to disk, there might be some interactions in system calls, which we do not analyze.
Despite this imprecision, {\sf ConfigCrusher} was able to produce more accurate results than the other approaches.
We hope to overcome this issue by analyzing system calls to obtain even more accurate results.\looseness=-1

\begin{framed}
	\noindent{\textbf{RQ1:} 
	\noindent {\sf ConfigCrusher}'s prediction error is statistically indistinguishable or lower than other approaches. {\sf ConfigCrusher}'s high accuracy is usually achieved with lower cost compared to the other accurate approaches.
	\looseness=-1
	}
\end{framed}

\subsection{RQ2: Instrumentation Overhead}
\label{rq2-overhead}
As explained in Sec.~\ref{instrument}, we observed excessive overhead when executing our instrumented systems.
With RQ2, we investigate how much overhead is induced by instrumenting regions.
To answer this question, we compared the instrumentation overhead and execution times of uninstrumented systems with systems instrumented with the unoptimized Algorithm~\ref{region-location-algorithm} and with systems instrumented with the optimized Algorithm~\ref{region-location-algorithm} and the propagation Algorithms~\ref{propagate-down-algorithm} and~\ref{propagate-up-algorithm}.

\paragraph{Procedure:}
We used the execution time of the uninstrumented systems as ground truth and executed the configuration that triggered the most number of regions in the optimized systems.
We executed the configuration with the highest execution time in case of multiple configurations with the same number of executed regions.\looseness=-1

\paragraph{Static and Dynamic Overhead Metric.} We measured the number of instrumented regions as the static overhead and the number of times the regions were entered and exited as the dynamic overhead. \looseness=-1

\paragraph{Time Metric.} We measured the execution times of the unoptimized and optimized instrumentations.\looseness=-1

\begin{table}[t]
\footnotesize
\centering
\caption{Static and dynamic comparison of instrumented regions before and after optimization for the configuration with the largest dynamic overhead.}
\begin{threeparttable}
\begin{tabular}{rrrrrrrrrr}
\toprule
\multicolumn{1}{c}{} & \multicolumn{1}{c}{Original} & \multicolumn{1}{c}{} & \multicolumn{3}{c}{Unoptimized} & \multicolumn{1}{c}{} & \multicolumn{3}{c}{Optimized} \\ \cline{2-2} \cline{4-6} \cline{8-10} 
S & Time &  & R & RC & Time &  & R & RC & Time \\ \midrule
1 & 12.14s &  & 16 & 32 & 12.16s &  & 10 & 22 & 12.13s \\
2 & 5.40s &  & 36 & $>10^6$ & $>$1h &  & 13 & 18 & 5.54s \\
3 & 3.68m &  & 397 & $>10^6$ & $>$1h &  & 7 & 88 & 3.77m \\
5 & 22.87s &  & 46 & $>10^6$ & $>$1h &  & 1 & 2 & 21.87s \\
6 & 1.04m &  & 128 & $>10^9$ & $>$1h &  & 23 & 4160 & 1.04m \\
7 & 26.08s &  & 60 & 8530 & 25.50s &  & 11 & 1204 & 25.49s \\
8 & 1.25m &  & 147 & $>10^9$ & $>$1h &  & 28 & $>10^6$ & 1.27m \\
9 & 4.87m &  & 166 & $>10^9$ & $>$1h &  & 1 & 2 & 4.89m \\
10 & 6.45m &  & 202 & 2420 & 6.53m &  & 10 & 78 & 6.45m \\
\bottomrule
\end{tabular}
S = Subject system;
R = \# of instrumented regions;
RC = \# of executed regions.
\end{threeparttable}

\label{region-reduce}
\end{table}

\paragraph{Results:}
Table~\ref{region-reduce} shows the results of our analysis. {\sf ConfigCrusher}'s optimized instrumentation with Algorithm~\ref{region-location-algorithm}, ~\ref{propagate-down-algorithm}, and~\ref{propagate-up-algorithm} reduced the number of regions and overhead by several orders of magnitude. By contrast, the unoptimized instrumentation, Algorithm~\ref{region-location-algorithm}, created an excessive amount of overhead, preventing running systems in a reasonable amount of time.\looseness=-1

Only the systems Running Example~($1$), Email~($7$), and Density Converter~($10$) had a low unoptimized dynamic overhead and executed in similar time compared to the original systems.
These systems do not have deeply nested regions, which does not increase the overhead of the dynamic analysis.
Prevayler~($8$), which has similar structure to the previous systems, executed a large number of optimized instrumented regions with low overhead.\looseness=-1

We can attribute the lower execution times of the instrumented systems compared to the original systems of the Running Example~($1$), Grep~($5$), and Email~($7$) to measurement noise.\looseness=-1

\begin{framed}
	\noindent{\textbf{RQ2:} 
	\noindent {\sf ConfigCrusher}'s optimized instrumentation incurs orders of magnitude less overhead compared to the unoptimized instrumentation.
	\looseness=-1
	}
\end{framed}

\subsection{RQ3: Performance Influence of Options in Regions}
\label{rq3-local}
One of the benefits of {\sf ConfigCrusher} over black-box approaches is that it builds local performance-influence models, which indicate whether and how options locally influence the performance of a system.
With RQ3, we investigate to how many regions the influence of options on performance can be localized.
To answer this question, we analyzed all local performance-influence models to determine the local influence of options on performance.
Subsequently, we manually examined the source code regions corresponding to the local models to further understand how they are influenced by options.
We conjecture that this type of information, derived from the local models, can provide insights for enhanced analysis of individual components of a system.\looseness=-1

\paragraph{Procedure: } 
We classified all local performance-influence models into two categories according to how options influence their performance.
Then, we manually analyzed all corresponding source code regions to understand how they are influenced by options.
For example, we classified the following models from Pngtastic Optimizer~($3$):
\begin{itemize}[noitemsep,topsep=0pt]
\item $\Pi_{\texttt{9cb}}$ = \verb|0.0|
\item $\Pi_{\texttt{b15}}$ = \verb|1.5|
\item $\Pi_{\texttt{b1a}}$ = \verb|0.2 + 82.5Compress + 142.1CompressIter + ...|,
\end{itemize}
\noindent
as "Not influenced by options" (e.g., $\Pi_{\texttt{9cb}}$ and $\Pi_{\texttt{b15}}$) and "Influenced by options" (e.g., $\Pi_{\texttt{b1a}}$).

\paragraph{Results: }
Table~\ref{region-stats} shows the results of our analysis. {\sf ConfigCrusher} helped us to identify that the influence of options on performance can be localized to a few regions in a system.
In these regions only subsets of all options interact.
The local performance-influence models helped us to easily locate these regions in the source code to further analyze this performance behavior.
In all such regions (e.g., Region~\texttt{b1a} above), the options influenced a loop or a control-flow statement within a loop, which either manipulated data structures or performed I/O operations.
These structures were sometimes located in the method where the region was instrumented, but other times we performed manual probing to find them in other methods called from the instrumented method.\looseness=-1

\begin{table}[t]
\footnotesize
\centering
\caption{Influence of options analysis of local performance-influence models.}
\begin{threeparttable}
\begin{tabular}{rrrrrrrrl}
\toprule
\multicolumn{1}{l}{} & \multicolumn{3}{c}{PNIO} & \multicolumn{1}{l}{} & \multicolumn{4}{c}{\multirow{2}{*}{ Performance influenced by options}} \\ \cline{2-4}
\multicolumn{1}{c}{} & \multicolumn{1}{c}{NEG} & \multicolumn{1}{c}{} & \multicolumn{1}{c}{Non-NEG} & \multicolumn{1}{c}{} & \multicolumn{4}{c}{} \\ \cline{2-2} \cline{4-4} \cline{6-9} 
S & Regions &  & Regions &  & Regions & Min ID & Max ID & Structure \\ \midrule
1 & 0 &  & 0 &  & 10 & 1 & 3 & Sleep \\
2 & 13 &  & 2 &  & 0 & N/A & N/A & Loop, I/O \\
3 & 3 &  & 1 &  & 3 & 3 & 3 & Loop, I/O \\
5 & 0 &  & 0 &  & 1 & 6 & 6 & Loop \\
6 & 19 &  & 2 &  & 2 & 6 & 6 & Loop, I/O \\
7 & 7 &  & 0 &  & 4 & 2 & 8 & Loop, Sleep \\
8 & 22 &  & 1 &  & 5 & 2 & 5 & Loop, I/O \\
9 & 0 &  & 0 &  & 1 & 8 & 8 & Loop \\
10 & 8 &  & 1 &  & 1 & 8 & 8 & Loop, I/O \\
\bottomrule
\end{tabular}
PNIO = Performance not influenced by options;
NEG = Negligible execution time (region which contribute $< 5\%$ of the execution time of the system). ID = Interaction degree.
\end{threeparttable}
\label{region-stats}
\end{table}

\paragraph{Local models discussion.} 
The local performance-influence models indicate the options that interact in the corresponding regions and whether and how they locally influence the performance of the system.
The exploratory analysis of the corresponding source code regions yielded some interesting findings of how options are implemented in these systems, which cannot be found with black-box approaches.

In a few regions, the control-flow statements with non-negligible execution time depended on options, yet the same branch of the control-flow statement was executed for all configurations.
This behavior was surprising since we selected, based on the systems' documentation, configuration values for each option that should behave differently.
In fact, we discovered that two options of Pngtastic Counter~($2$) were not used in the source code as they were described in the documentation.
For example, the valid range of an option was $0.0 - 1.0$, and we conjecture that the system would behave differently by picking different values.
However, the control-flow statement where this option was used always executed the same branch if the value was $>0$.
Finding these inconsistensies, common in configurable systems~\citep{XZHZSYZP:SOSP13, RK:ICSE11, HY:ESEM16, CCRC:ASE18}, might be useful for developers and maintainers to debug these type of systems.\looseness=-1

Interestingly, some options influenced only regions with negligible execution time.
This behavior was surprising since we expected, based on the systems' documentation, that the options would influence the performance of the systems; for example, how the options that influence Region \texttt{9cb} above drastically change its execution time.
We manually confirmed that they involve either a few statements or did not contain expensive loops nor calls.
While black-box approaches also found that these options do not influence the performance, {\sf ConfigCrusher} helped us to pinpoint the regions where these options are used to understand this potentially unexpected behavior.\looseness=-1



We conjecture that developers can potentially discern similar findings in other configurable systems to make more informative decisions during debugging and optimization of these systems.\looseness=-1

\begin{framed}
	\noindent{\textbf{RQ3:} 
	\noindent {\sf ConfigCrusher} helped us to identify that the influence of options on performance is localized to a few regions in a system.
The options influence loops or control-flow statements within loops.
	\looseness=-1
	}
\end{framed}

\subsection{Threats to Validity}
\label{threats}
The primary threat to external validity is the selection of subject systems. 
As discussed in Sec.~\ref{subject-systems}, we were limited by the overhead and precision of the static analysis.
This limitation is due to the novelty of our approach and shared with other white-box approaches, though we lift some of their limitations.
While we selected a set of widely-used open-source Java configurable systems from different domains, readers should be careful when generalizing results to other types of systems. 
For instance, we analyzed single-threaded systems with deterministic execution time. 
Additionally, we analyzed and built a performance-influence model for a single configurable system.  
Systems composed of numerous configurable systems, deployed in distributed environments, and implemented in different languages are beyond the scope of this article.

Another threat to validity is the subset of options that we selected for analysis in the subject systems. 
We selected options for which the systems’ documentation or the options’ functionality indicated that they would affect performance and observed a wide range of execution times for the configurations that we measured.

Another threat to validity is the selection of the data-flow analysis.
As discussed in Sec.~\ref{implementation}, we selected the state of the art static taint analysis, but reduced its precision in favor for an analysis that terminates.
This strategy has been used in previous work and, as demonstrated in our evaluation, our approach is robust and produces accurate results with the settings that we selected.\looseness=-1

Measurement noise cannot be excluded and may affect the results
we obtained. 
We reduced this threat by repeating measurements
several times on a dedicated machine and averaging the results.

\section{Related Work}
In Sec.~\ref{state-of-the-art}, we described performance modeling in general and closely related state of the art black-box and white-box approach for performance-influence modeling and compared them to {\sf ConfigCrusher}.
In this section, we discuss additional research to position {\sf ConfigCrusher} and white-box analyses in the context of prior work.\looseness=-1

\emph{Analysis of configurable systems:} 
Similar to our work, several researchers have leveraged some kind of program analysis to explore various properties of configuration options~\citep{HSCMAR:ASPLOS11, NKCFP:FSE16, DALC:ISSRE16, RK:ICSE11, WPXCZZ:SPLC13, SD:JSS18, RSMFP:ICSE10, MWKTS:ASE16}.
Th\"{u}m et al.~\citep{TAKSS:CSUR14} presented a comprehensive survey of analyses for software product lines also applicable to configurable systems.\looseness=-1

Similar to our approach, Lillack et al.~\citep{LKB:TSE18} used taint analysis to identify, for each code fragment, in which configurations it may be executed.
However, they do not track information about individual options.
Instead, our taint analysis tracks how options influence code fragments due to control-flow and data-flow interactions to track how options influence the performance of the system.\looseness=-1

Hoffmann et al.~\citep{HSCMAR:ASPLOS11} used dynamic influence tracing to convert static parameters into dynamic control variables to adapt properties of an application.
However, they do not consider interactions beyond two parameters.
By contrast, our approach tracks control-flow and data-flow interactions among options, without a limit on the interaction degree, and how they influence the performance of the system.\looseness=-1

Reisner et al.~\citep{RSMFP:ICSE10} and Meinicke et al.~\citep{MWKTS:ASE16} used symbolic execution and variational execution, respectively, to analyze the behavior of interactions in configurable systems and found the insights of Irrelevance, Independence, and Low Interaction Degree that we consider in this work .
We leverage those insights to create a novel white-box performance analysis that efficiently generates accurate performance-influence models.\looseness=-1

\emph{Testing configurable systems:} Combinatorial Testing~\citep{KKL:ICT13, NL:CSUR11,HBG:SRE11,HMGB:IST16,HNADPB:ESE18} is an approach to reduce the number of samples to test a system by satisfying a certain coverage criterion.
Similarly, Souto et al.~\citep{SDG:ICSE17} improved SPLat~\citep{KMKBSBD:ESECFSE13} to use sampling heuristics~\citep{MKRGA:ICSE16} to select what configurations to sample.
While both these approaches scale to large systems, they make assumptions on how options interact in the system and can potentially miss relevant interactions.
Instead, our sampling is guided by white-box information on how options are used and interact in the systems.\looseness=-1

\emph{Performance profiling:} Several profiling techniques, including sampling and instrumentation, can be used to identify performance hot spots~\citep{MWX:ISSTA17, YP:EMSE18, CLBKMG:ICSECP18, G:CAMC16}.
For example, Castro et al.~\citep{CAPPJ:TACO15} used both techniques to identify hot spots that can be isolated and replayed as standalone systems for further performance analysis and optimization.
Our approach is complementary to this line of work, assisting in potentially narrowing down the performance-intensive components for more comprehensive profiling.

\emph{Energy measurement:} Modeling power or energy consumption is closely related to performance and employs similar techniques~\citep{JSBM:ISSTA16, GZBNBE:ESEM14, GLWH:GREENS16}.
For example, Hao et al.~\citep{HLHG:ICSE13} used program analysis to estimate the energy consumption of instructions of Android apps.
This line of work, however, does not address configurability, but could benefit from our approach to understand how the configuration of the system influences its energy consumption.\looseness=-1

\section{Conclusion}
This article presents {\sf ConfigCrusher}, a white-box performance analysis approach for configurable systems.
{\sf ConfigCrusher} employs a data-flow analysis to identify how configuration options may influence control-flow statements and instruments regions corresponding to those statements for performance measurement.
Our evaluation on $10$ real-word systems shows the potential of our white-box approach to builds similar or more accurate performance-influence models than other approaches with lower cost.
In contrast to state of the art approaches, our white-box approach provides additional information of the components of a system, which can aid stakeholders to analyze, optimize, and debug them.
Our work can provide a foundation for future research on white-box analysis for configurable systems, such as understanding the performance of larger and distributed systems, explaining causes for performance differences,  and combining white-box and black-box approaches for more accurate and efficient performance analysis.
\looseness=-1

\section{Acknowledgments}
This work has been supported in part by the NSF (awards 1318808, 1552944, and 1717022), AFRL, DARPA (FA8750-16-2-0042), and the German Research Foundation (AP 206/7-2, AP 206/11-1, SI 2171/2, SI 2171/3-1). We thank Chu-Pan Wong and Jens Meinicke for their comments during the development of this work. We thank the FOSD 2017 and 2018 meeting participants for their feedback on the central idea of this work. We thank Steven Artz for his help with FlowDroid.



%
%

\bibliographystyle{spbasic}      
\bibliography{bibliography.bib}

\end{document}